\documentclass[trackchanges]{aastex701}
\usepackage{graphicx}
\usepackage{bm}
\usepackage{color, soul}      
\usepackage{units}
\usepackage{amsmath}
\usepackage{epsfig}
\usepackage{amssymb}
\usepackage{mathrsfs}
\usepackage{comment}
\usepackage{hyperref}
\hypersetup{linkcolor=red,citecolor=blue,filecolor=cyan,urlcolor=blue}

\renewcommand{\vec}[1]{\boldsymbol{#1}}

\begin{document}

\title{
Energy partition in resonant instabilities in multi-component plasmas
}

\author[0000-0003-3223-1498]{Rodrigo A. L\'opez}
\affiliation{Research Center in the intersection of Plasma Physics, Matter, and Complexity ($P^2 mc$), Comisi\'on Chilena de Energ\'{\i}a Nuclear, Casilla 188-D, Santiago, Chile}
\affiliation{Departamento de Ciencias F\'{\i}sicas, Facultad de Ciencias Exactas, Universidad Andres Bello, Sazi\'e 2212, Santiago 8370136, Chile}
\email[show]{rodrigo.lopez@cchen.cl}

\author[0000-0001-6038-1923]{Kristopher G. Klein}
\affiliation{Lunar and Planetary Laboratory, University of Arizona, Tucson, AZ, USA}
\email{}

\author[0000-0002-8820-8177]{Francisco Ley}
\affiliation{Departamento de F\'isica, Facultad de Ciencias F\'isicas y Matem\'aticas, Universidad de Chile, Santiago 8370448, Chile}
\email{}

\author[0000-0002-9161-0888]{Pablo S. Moya}
\affiliation{Departamento de F\'isica, Facultad de Ciencias, Universidad de Chile, Las Palmeras 3425, Santiago, 7800003, Chile}
\email{pablo.moya@uchile.cl}

\author[0000-0003-3483-4890]{Aaron Tran}
\affiliation{Department of Physics, University of Wisconsin–Madison, 1150 University Ave, Madison, WI 53706, USA}
\email{atran@physics.wisc.edu}

\author[0000-0002-0497-1096]{Daniel Verscharen}
\affiliation{Mullard Space Science Laboratory, University College London, Dorking, RH5~6NT, United Kingdom}
\affiliation{Ruhr University Bochum, Faculty of Physics and Astronomy, Theoretical Physics I, Theoretical Plasma Physics, 44801 Bochum, Germany}
\email{daniel.verscharen@rub.de}  

\author[0000-0003-4887-9512]{Nicol\'as Villarroel-Sep\'ulveda}
\affiliation{Departamento de F\'isica, Facultad de Ciencias F\'isicas y Matem\'aticas, Universidad de Concepci\'on, Concepci\'on 4070386, Chile}
\email{}

\begin{abstract}
Space and astrophysical plasmas are often collisionless and exhibit non-equilibrium velocity distributions. When these departures from equilibrium are sufficiently large, the plasma becomes unstable to various kinetic micro-instabilities, which provide a mechanism for transferring energy from the particles to the electromagnetic (EM) fields. In multi-component plasmas, different species interact with the generated EM waves in distinct ways, acting as sources or sinks of wave energy through resonant wave--particle interactions. Consequently, the nonlinear evolution of kinetic instabilities is governed not only by the species driving the instability but also by the resonant response of other plasma populations. In this work, considering proton-$\alpha$-particle plasmas with parameters representative of the solar wind, we investigate the generation, saturation, and energy partition of kinetic micro-instabilities, such as Alfv\'en/ion-cyclotron and fast-magnetosonic/whistler wave instabilities. We combine linear Vlasov--Maxwell theory and hybrid particle-in-cell (PIC) simulations to analyze these instabilities and to quantify the exchange of energy and momentum between ion populations during the instability evolution. Our results demonstrate that minor ions play a fundamental role in regulating the nonlinear saturation of kinetic instabilities and the transfer of energy among plasma populations. 
\end{abstract}



\section{Introduction}
\label{sec:intro}

Space and astrophysical plasmas are often in a collisionless state, and many natural plasma systems exhibit non-equilibrium particle velocity distribution functions \citep[VDFs;][]{Marsch2006,Verscharen2019}. 
For example, space plasmas like the solar wind often exhibit temperature anisotropies with respect to the local magnetic field and field-aligned relative drifts between plasma species \citep{Hundhausen1967,Asbridge1976,Marsch1982}. Additional deviations from the Maxwell--Boltzmann distribution persist, for instance, in the form of suprathermal tails or more localized deviations in velocity space \citep{Goodrich1976,Christon1988,Espinoza2018,Eyelade2025}.

If these deviations from equilibrium are large, the plasma becomes unstable and excites kinetic micro-instabilities \citep{Gary1993}. During the linear phase of these instabilities, the amplitudes of the unstable normal modes of the plasma grow exponentially in time at the expense of the free energy contained in the driving non-equilibrium features in the particle VDFs. 
In the case of resonant instabilities, particle scattering on the growing fluctuations in the electromagnetic field of the unstable wave modifies the particle energy in such a way that the total energy of the system remains conserved \citep{Kennel1966,Kennel1967}. Micro-instabilities are thus a mechanism for transferring energy from plasma particles to the electromagnetic field.

In the case of the solar wind, the dynamically most important plasma species are the protons, the electrons, and the $\alpha$-particles. 
All of these species exhibit deviations from Maxwellian equilibrium distributions. 
For instance, the protons often show a core ($j=\mathrm c$) and a beam ($j=\mathrm b$) population, each of which presents different bulk velocities $\vec U_j$ and temperature anisotropies with $T_{\perp j}\ne T_{\parallel j}$, where $T_{\perp j}$ and $T_{\parallel j}$ are the temperatures of species $j$ in the directions perpendicular and parallel with respect to the magnetic field \citep{Feldman1973,DeMarco2023,Bruno2024}. 
Moreover, the $\alpha$-particles often have a bulk speed $\vec U_{\alpha}\ne \vec U_{\mathrm p}$ \citep{Marsch1982b,Reisenfeld2001}. 
Ultimately, all of these deviations from equilibrium are potential drivers of kinetic micro-instabilities \citep{Montgomery1976,Marsch1987,Gary1991,Daughton1998,Gary2000,Goldstein2000,Hellinger2006,Matteini2007,Bale2009,Podesta2011,Maruca2012,Bourouaine2013,Verscharen2013,Jian2016,Wicks2016,Klein2018,Navarro2020,Verniero2020}.

The response of different plasma species $j$ to the unstable normal modes depends on the individual VDFs $f_j$ as well as the charge $q_j$ and mass $m_j$ of the species. 
In such a multi-species plasma, some species may release energy while others absorb energy through resonant wave--particle interactions. An overall micro-instability with net energy transfer from the plasma particles to the electromagnetic field is possible only if at least one species releases more energy than the combined energy absorption of the other species.

While some species emit and others absorb energy, their VDFs $f_j$ change through pitch-angle scattering, modifying the velocity moments of the VDF according to the kinetic theory of resonant wave--particle interactions \citep{Rowlands1966,Shapiro1966,Tu2002}. Micro-instability can thus couple all involved particle species and cause free energy to flow from one driving species into other species via electromagnetic fluctuations.
In the following, we quantify the energetics of this inter-species coupling via kinetic micro-instabilities in multi-species plasmas.

Here, using linear Vlasov--Maxwell theory and kinetic plasma simulations, we demonstrate inter-species coupling via kinetic micro-instabilities in plasma configurations comparable to those observed in the solar wind. This article is organized as follows. In Section~\ref{sec:energetics}, we develop quantitative arguments for the energetics of inter-species coupling through instabilities. In Section~\ref{sec:methodology}, we present the theory and methods we use to analyze wave--particle interactions and plasma energetics through kinetic theory and simulations, considering four Cases selected to excite kinetic instabilities. Then, Sections~\ref{sec:ICW} and \ref{sec:FMW} show the results considering the 
Alfvén/ion-cyclotron (A/IC) and fast-magnetosonic/whistler (FM/W) instabilities. Finally, in Section~\ref{sec:discussion}, we discuss our results and summarize our main conclusions.

\section{Energetics of the inter-species coupling through instabilities}\label{sec:energetics}

In a plasma consisting of particles of species $j$, the total energy is conserved
\footnote{The total energy in Equation~(\ref{Wtot}) is strictly conserved only for closed systems. Natural plasmas almost always have open boundaries with non-zero particle, momentum, and energy fluxes across any finite volume boundaries. These fluxes require a more careful treatment of the concept of energy conservation. However, our numerical simulations here use periodic boundary conditions, allowing us to make the assumption of a closed volume for the sake of energy conservation.}:
\begin{equation}\label{Wtot}
W=\sum\limits_jW_j+W_{EB}=\text{constant},
\end{equation}
where
\begin{equation}
W_j=\frac{1}{2}m_j\iint v^2f_j\,\mathrm d^3 v\,\mathrm d^3x,
\end{equation}
is the kinetic energy of the particles of species $j$. We define the electromagnetic energy $W_{EB}=W_E+W_B$, where
\begin{equation}
W_E=\frac{1}{8\pi}\int E^2\,\mathrm d^3x
\end{equation}
is the energy in the electric field $\vec E$\footnote{We demonstrate our work through simulations with a non-relativistic hybrid code. In such a code, the electric field transfers energy but is not a reservoir. In the assumed limit of these simulations, for which $c\rightarrow \infty$, the energy in the electric field does not contribute to the overall energy conservation and $W_{EB}\approx W_B$.}, and
\begin{equation}
W_B=\frac{1}{8\pi}\int B^2\,\mathrm d^3x
\end{equation}
is the energy in the magnetic field $\vec B$. The integration over $\mathrm d^3v$ is to be taken over the entire velocity space, and the integration over $\mathrm d^3x$ over the entire volume of the plasma.

With this notation, we characterize instabilities by a net increase in $W_{EB}$ at the expense of a reduction in $\sum_jW_j$ over time, while the overall $W$ remains constant. A net increase in $W_{EB}$ can be achieved even when one or more species gain energy, i.e., their individual $\delta W_j>0$, as long as the sum over all $\delta W_j$ is negative. 

The arguments above based on the total energy balance according to Equation~(\ref{Wtot}) are sufficient for the description of the energetics of instabilities and, by extension, to cases with a net \emph{decrease} in $W_{EB}$, of wave damping. However, for a closer inspection of the energy transfer involved, it is useful to separate $W_j$ into contributions from the bulk kinetic energy and the thermal energy of species $j$. We assume that $f_j$ is gyrotropic. With the definitions of the number density
\begin{equation}
n_j=\int f_j\,\mathrm d^3v
\end{equation}
and the bulk velocity
\begin{equation}
\vec U_j=\frac{1}{n_j}\int \vec v f_j\,\mathrm d^3v
\end{equation}
of species $j$, we then rewrite $W_j$ as
\begin{equation}\label{Wjsplit}
W_{j}=W_{Uj}+W_{\mathrm {th}j},
\end{equation}
where 
\begin{equation}
W_{Uj}=\frac{1}{2}m_j \int n_jU_j^2\,\mathrm d^3x,
\end{equation}
is the bulk kinetic energy contribution for species $j$,  
\begin{equation}
W_{\mathrm {th}j}=\int\left(n_jk_BT_{\perp j}+\frac{1}{2} n_jk_BT_{\parallel j}\right)\,\mathrm d^3x,
\end{equation}
the thermal energy contribution of species $j$, and $k_B$ is the Boltzmann constant. In this derivation, we use the definition of the pressure tensor
\begin{equation}
\mathsf P_{j}=m_j\int \left(\vec v-\vec U_j\right) \left(\vec v-\vec U_j\right) f_j\,\mathrm d^3v
\end{equation}
and define the kinetic temperatures of species $j$ perpendicular and parallel to the magnetic field as
\begin{equation}
T_{\perp j}=\frac{\mathrm{Tr}\left[\mathsf P_j\left(\mathbf 1-\hat{\vec b}\hat{\vec b}\right)\right]}{2n_jk_B}
\end{equation}
and
\begin{equation}
T_{\parallel j}=\frac{\mathsf P_j\hat{\vec b}\hat{\vec b}}{n_jk_B},
\end{equation}
where $\hat{\vec b}=\vec B/B$ is the unit vector of the magnetic field.

Equation~(\ref{Wjsplit}) separates the particle kinetic energy of species $j$ into the contributions from the bulk kinetic energy and the thermal energy. This separation allows us to determine whether, for a given instability, the bulk motion or the thermal pressure is the main source or sink of particle energy for a given species $j$. If there is a meaningful distinction between multiple populations of particles of the same species, e.g., proton core and proton beam populations, the framework also applies and allows for them to be separated as different $j$.

\section{Methodology}
\label{sec:methodology}

To investigate the redistribution of energy among ion populations by instability-driven fluctuations, we consider a set of drifting multi-species plasma configurations representative of conditions in the solar wind. We perform linear predictions and non-linear simulations for each configuration. The selected parameters are chosen to excite either Alfvén/ion-cyclotron (A/IC) or fast-magnetosonic/whistler (FM/W) instabilities, while simultaneously allowing substantial resonant interactions with non-driving ion species and thus energy transfer between species through the excited electromagnetic fluctuations. These configurations are designed such that $\delta W_j$ is positive for one species but negative for another, while $\delta W_{EB}$ is positive.

\begin{figure}
    \centering
    \includegraphics[width=0.8\linewidth]{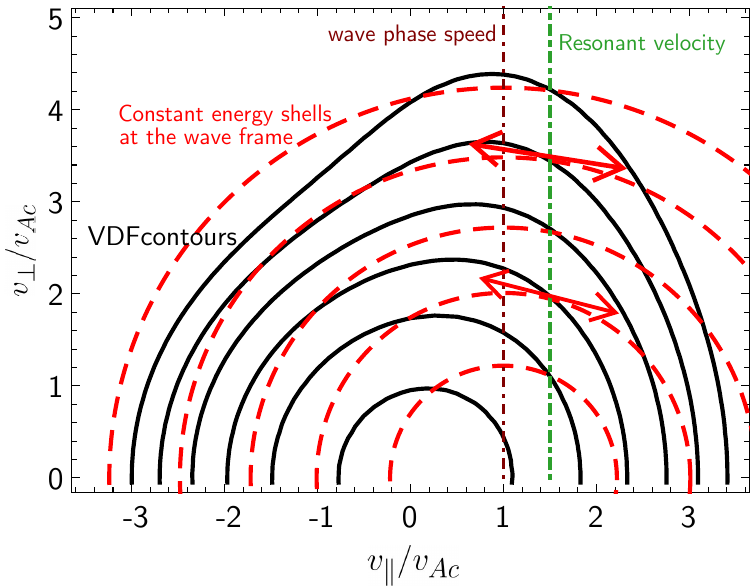}
    \caption{Sketch of quasilinear diffusion due to resonant interactions in ($v_\perp$, $v_\parallel$) space. Solid black and dashed red curves represent constant VDF contours and the constant energy shells in the wave reference frame, respectively. Dot-dashed dark red and green vertical lines correspond to the field-parallel component of the phase velocity of the wave and the field-parallel velocity of cyclotron-resonant particles. Red arrows indicate the directions of the diffusive flux of resonant particles through velocity space.}
    \label{fig:cartoon}
\end{figure}

According to quasilinear theory, as illustrated in Fig.~\ref{fig:cartoon}, wave--particle interactions with the emitted waves cause cyclotron-resonant particles (represented by the vertical green line) to pitch-angle scatter and subsequently diffuse locally in the direction of the tangents (red arrows) of circles given by $v^2_\perp + (v_\parallel - \omega_{\mathrm r}/k_\parallel)^2 = \rm{constant}$, where $v_{\perp}$ and $v_{\parallel}$ are the velocity components perpendicular and parallel to the local background field. These circles represent shells of constant energy in the frame comoving with the field-parallel component of the wave's phase velocity $\omega_{\mathrm r}/k_{\parallel}$ \citep{Yoon2017,Verscharen2019}, where $\omega_{\mathrm r}$ is the real part of the wave frequency and $\vec k$ is the wavevector.

We model the plasma as a superposition of bi-Maxwellian background VDFs:
\begin{equation}
    \label{eq:f0}
    f_{0j}(v_\perp,v_\parallel) = \frac{n_j}{\pi^{3/2} w_{\perp j}^2 w_{\parallel j}}\,\exp\left(-\frac{v^2_\perp}{w^2_{\perp j}} - \frac{(v_\parallel - U_j)^2}{w^2_{\parallel j}} \right),
\end{equation}
where $w^2_{\perp j} = 2k_B T_{\perp j}/m_j$ and $w^2_{\parallel j} = 2k_B T_{\parallel j}/m_j$ are the squares of the perpendicular and parallel thermal speeds of the species $j$. 

We study four Cases, detailed in Table~\ref{t1}. 
In all Cases, densities are normalized to the proton (core) density $n_\text{c}$, the ratio of parallel thermal to magnetic pressure for each $j$ is defined as $\beta_{\parallel j}=8\pi n_jk_BT_{\parallel j}/B_0^2$, and the drift speed is normalized to the proton (core) Alfv\'en speed $v_\text{Ac}=B_0/\sqrt{4\pi n_\text{c}m_\text{p}}$, where $\vec B_0$ is the background magnetic field, and $m_\text{p}$ the proton mass. 

Case 1 is composed of three populations: protons (c), drifting $\alpha$-particles $(\alpha)$, and electrons (e). This Case is designed to drive the A/IC instability through the anisotropic $\alpha$-particle population with $T_{\perp \alpha}>T_{\parallel \alpha}$ and a relative drift similar to the proton Alfv\'en speed $v_\text{Ac}$ with respect to the isotropic proton population. This system has $\delta W_\alpha<0$ and $\delta W_\text{c}>0$.

In Case 2, we include a proton beam, leaving the system composed of four species: a proton core (c), proton beam (b), drifting $\alpha$-particles $(\alpha)$, and electrons (e). In this Case, we study the proton-beam-driven FM/W instability, keeping all species isotropic and introducing a large beam drift (in units of the proton core Alfv\'en speed). The $\alpha$-particles, while also drifting, do not contribute to the driving of the instability but instead receive energy through wave--particle interactions with the proton-beam-driven unstable waves. This system has $\delta W_\text{b}<0$, $\delta W_\text{c}>0$ and $\delta W_\alpha>0$.

Finally, Cases 3 and 4 serve as control cases for Case 2. Here, we systematically reduce the $\alpha$-particle population concentration (to 1\% in Case 3 and 0\% in Case 4) to analyze the energy distribution between species when the concentration of a non-driving ion species decreases. These systems have $\delta W_\text{b}<0$ and $\delta W_\text{c}>0$,
while $\delta W_\alpha\ge0$ in Cases 3 and 4.

In all four Cases, the electron density and drift velocity with respect to the protons have been selected to guarantee quasi-neutrality and a zero parallel current condition.

\begin{deluxetable*}{l l c c c c}
\tablecaption{Plasma parameters for our four Cases under investigation.\label{tab:params}}
\tablewidth{0pt}
\tablehead{
\colhead{Case} &
\colhead{Species} &
\colhead{$n_j/n_\text{c}$} &
\colhead{$\beta_{\parallel j}$} &
\colhead{$T_{\perp j}/T_{\parallel j}$} &
\colhead{$U_j/v_\text{Ac}$}
}
\startdata
1 & Protons (c)       & 1.0  & 3.0  & 1.0 & 0.0   \\
  & $\alpha$-particles ($\alpha$) & 0.05 & 0.6  & 4.0 & 0.9   \\
  & Electrons (e)     & 1.1  & 3.3  & 1.0 & 0.082 \\
\hline
2 & Proton Core (c)   & 1.0  & 1.0  & 1.0 & 0.0   \\
  & Proton Beam (b)   & 0.2  & 0.2  & 1.0 & 1.8   \\
  & $\alpha$-particles ($\alpha$) & 0.05 & 0.3  & 1.0 & 0.8   \\
  & Electrons (e)     & 1.3  & 1.3  & 1.0 & 0.338 \\
\hline
3 & Proton Core (c)   & 1.0  & 1.0  & 1.0 & 0.0   \\
  & Proton Beam (b)   & 0.2  & 0.2  & 1.0 & 1.8   \\
  & $\alpha$-particles ($\alpha$) & 0.01 & 0.06 & 1.0 & 0.8   \\
  & Electrons (e)     & 1.22 & 1.22 & 1.0 & 0.308 \\
\hline
4 & Proton Core (c)   & 1.0  & 1.0  & 1.0 & 0.0   \\
  & Proton Beam (b)   & 0.2  & 0.2  & 1.0 & 1.8   \\
  & Electrons (e)     & 1.2  & 1.2  & 1.0 & 0.3   \\
\enddata
\tablecomments{Case 1 contains a drifting anisotropic $\alpha$-particle population driving the A/IC instability, while Cases 2--4 include a proton-beam component triggering the FM/W instability. Case 4 does not include $\alpha$-particles. For the linear theory, we have set the Alfv\'en speed to $v_\text{Ac}/c=10^{-4}$. }
\label{t1}
\end{deluxetable*}

\subsection{Linear Theory}

We perform a linear analysis of Vlasov--Maxwell theory for each Case shown in Table~\ref{t1} using the \texttt{PLUME}~\cite[Plasma in a Linear Uniform Magnetized Environment;][]{PLUME_2025, Klein_2025} dispersion-relation solver.
\texttt{PLUME} solves the hot-plasma dispersion relation detailed in Chapter 10 of \cite{Stix:1992} for a plasma system consisting of an arbitrary number of species with relatively drifting bi-Maxwellian VDFs.
The solutions require specifying a reference ratio for the thermal to magnetic pressures $\beta_{\parallel R}$ and a reference ratio for the thermal speed to the speed of light  $w_{\parallel R}/c$ for reference species $R$. 
For our calculations, we select the proton core as the reference species ($R=c$).
Each component $j$ is further specified by a set of parameters $\mathcal{P}$ for each species $j$:
the temperature anisotropy $T_{\perp j}/T_{\parallel j}$, the 
temperature disequilibrium $T_{\parallel c}/T_{\parallel j}$,
the relative density ratio $n_j/n_\text{c}$, the relative charge ratio $q_\text{p}/q_j$,
the relative mass ratio $m_\text{p}/m_j$, and the drift parallel to the mean magnetic field normalized to the reference Alfv\'en velocity $U_j/v_{A c}$. 

For a given background specified by $\mathcal{P}$, \texttt{PLUME} determines the complex normal-mode solutions $\omega(\vec{k},\mathcal{P})=\omega_{\rm r}+i \gamma$ for a specified wavevector $\vec{k}=(k_\perp,k_\parallel)$. Here, $\omega_{\rm r}$ is the real part of the frequency of the normal mode oscillation and $\gamma$ is its growth $(\gamma>0)$ or damping $(\gamma<0)$ rate.
\texttt{PLUME} can  follow  solutions along user-defined paths in $\vec{k}$ and/or $\mathcal{P}$. For this work, we focus on variations in $k_\parallel$ with fixed $k_\perp$ and fixed $\mathcal{P}$ for a given Case. If $\gamma>0$, the wave amplitude grows with time, meaning that $\delta W_{EB}>0$. Likewise, if $\gamma<0$, the wave amplitude decreases and $\delta W_{EB}<0$.

In addition to determining the normal-mode solutions, \texttt{PLUME} calculates the eigenfluctuations of the density, velocity, and electromagnetic fields \citep{Klein_2025}, as well as of the fluctuations of the VDF itself \citep{Brown:2026}. The code also determines the collisionless power absorption or emission from each component:
\begin{equation}
P_j = \frac{\omega}{8 \pi} \vec{E}^* \cdot \underline{\underline{\chi}}_{j,a}\cdot \vec{E},
\label{eqn:power.hp}
\end{equation}
where according to  \citet{Stix:1992}, $\underline{\underline{\chi}}_{j,a}=
(\underline{\underline{\chi}}_j-\underline{\underline{\chi}}_j^\dagger)/2i$ is the anti-Hermitian susceptibility tensor for component $j$ with
$\underline{\underline{\chi}}_j^\dagger \equiv [\underline{\underline{\chi}}_j^\top]^*$
evaluated at $\omega = \omega_{\rm r}$.
From $P_j$, \texttt{PLUME} calculates the contributions $\gamma_j$ of species $j$ to the overall $\gamma$, such that $\sum_j\gamma_j=\gamma$ if $|\gamma/\omega_{\mathrm r}|\ll 1$. If $\gamma_j>0$, species $j$ makes a positive contribution to the overall $\gamma$, meaning that $\delta W_j<0$. Likewise,  if $\gamma_j<0$, species $j$ makes a negative contribution to the overall $\gamma$ and $\delta W_j>0$.

The code further decomposes this component emission or absorption into contributions from the different resonance orders $n$, which are defined in the resonance condition
\begin{equation}\label{eq:rescond}
\omega_{\mathrm r}=k_{\parallel}v_{\parallel}+n\Omega_j.
\end{equation}
The contributions $\gamma_j^n$ for different $n$ fulfill $\sum_{n}\gamma_j^n=\gamma_j$ if $|\gamma/\omega_{\mathrm r}|\ll 1$. 

The full expressions are given by \citet{Klein_2025}. For this work, we focus on the orders $n=0$ (Landau and transit-time resonances) and $n=\pm 1$ (cyclotron resonances). An example of this decomposition is shown in rows 3 and 4 of Fig.~\ref{fig:plume_case1}.

\subsection{Simulations}

For our nonlinear study, we employ one-dimensional, non-relativistic ($c\to\infty$) hybrid particle-in-cell (PIC) simulations for each Case, in which ions are modeled as kinetic particles and electrons as a charge-neutralizing fluid. We use the code Hybrid-VPIC~\citep{Bowers2008, Le2021,Le2023}, a high-performance open-source hybrid PIC code \footnote{available at \href{https://github.com/lanl/vpic-kokkos/tree/hybridVPIC}{https://github.com/lanl/vpic-kokkos/tree/hybridVPIC}
(Git commit \texttt{08f2a23c}).
}

The one-dimensional setup is built by $N_x=2048$ cells in a box of length $L_x=512\,d_\text{c}$ oriented along the $x$-direction, and $\vec B_0=B_0\hat{\vec e}_x$,  where $d_\text{c}=v_{Ac}/\Omega_{p}$ is the proton (core) inertial length. We initialize the system with $n_\text{ppc}=100,000$ particles per cell per component (core and beam protons, and $\alpha$-particles). We set the time-step to be $\Delta t=0.02\,/\Omega_{p}$, with the simulation running until $T_\text{max}=655.36\,/\Omega_{p}$. We use a number of particles per cell that ensures energy conservation (see Appendix~\ref{sec:appendix_conservation}). The electron fluid is isothermal with scalar pressure. We don't include explicit collisions or dissipation terms, and we set resistivity and hyper-resistivity to zero.

We scale the numerical particle weights for beam protons and $\alpha$-particles to their density ratios $n_\text{b}/n_\text{c}$ and $n_\alpha/n_\text{c}$. We use a quadratic-sum particle shape to deposit the current and charge density on the Cartesian mesh \citep[Appendix~B]{Le2023}; Also, we apply three passes of a ``1-2-1'' binomial filter to the current and charge on the mesh, and also to the electromagnetic fields when interpolating to particles.

\section{Alfv\'en Ion-cyclotron instability}
\label{sec:ICW}

We begin our analysis with Case 1 in Table~\ref{tab:params}, representing a plasma in which the free energy to drive the instability is carried by the $\alpha$-particles with a temperature anisotropy $T_{\perp \alpha}/T_{\parallel \alpha} = 4$ and drift $U_{\alpha} = 0.9$~$v_\text{Ac}$. The resulting shape of their VDF makes the $\alpha$-particles susceptible to the resonant driving of A/IC instability. Next, we perform a linear analysis, evaluate the dispersion relation to identify unstable modes, and determine each ion species' contribution to the overall growth rate. We then present results from our PIC simulations and compare the linear predictions with the plasma's nonlinear evolution. 

\subsection{Linear Prediction}
\label{ssec:aic.linear}


Figure~\ref{fig:plume_case1} shows the linear analysis for Case 1 for the A/IC instability. The panels show, from top to bottom, the normalized real frequency $|\omega_{\mathrm r}|$, growth/damping rate $|\gamma|$, proton contribution to the growth rate $|\gamma_\text{c}^{n}|$, and $\alpha$-particle contribution $|\gamma_\alpha^{n}|$ as functions of parallel wavenumber $k_\parallel d_\text{c}$. Solid and dashed curves denote $\gamma>0$ and $\gamma<0$, respectively. We identify four parallel-propagating modes: two FM/W (not shown) and two A/IC modes.
The FM/W modes are stable over the investigated wavevector range. The second row shows the unstable behavior of the forward- and backward-propagating left-handed modes, A/IC A and A/IC B, respectively, consistent with the properties of the A/IC instability. 
This finding is supported by the wavevector-dependence of the real part of the frequency (first row) and the transverse polarization (not shown). The third and fourth rows give us information about the power absorption/emission of each species, showing that the unstable modes are driven by the $\alpha$-particles that cyclotron-resonate with $n=\pm1$, while they are subject to significant proton-cyclotron damping via the same resonances. This combined behavior shows a direct pathway for energy transfer from $\alpha$-particles to protons, mediated by A/IC waves.

\subsection{Simulation Results}
\label{ssec:aic.simulation}

\begin{figure}
    \centering
    \includegraphics[width=0.8\linewidth]{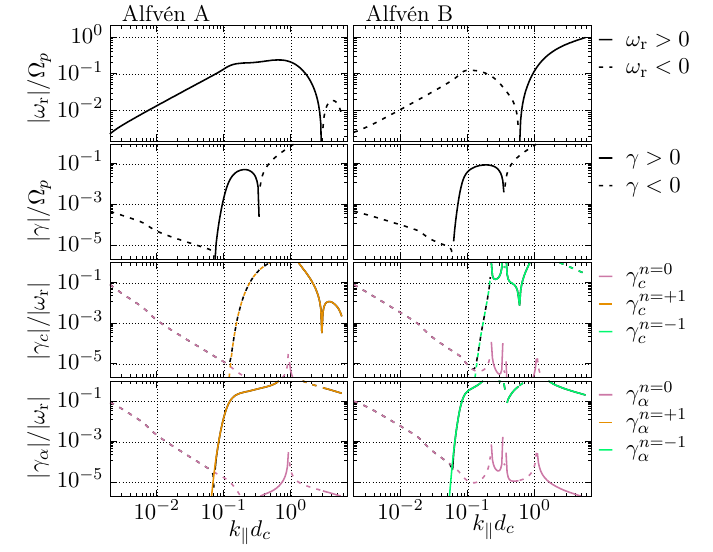}
    \caption{Linear dispersion analysis for Case 1, corresponding to an $\alpha$-particle-driven A/IC instability. From top to bottom, the panels show the normalized real frequency $|\omega_{\mathrm r}|$, growth/damping rate $|\gamma|$, proton contribution to the growth rate $|\gamma_\text{c}^{n}|$, and $\alpha$-particle contribution $|\gamma_\alpha^{n}|$, as functions of parallel wavenumber $k_\parallel d_\text{c}$. In the first row, solid and dashed curves denote $\omega_{\mathrm r}>0$ and $\omega_{\mathrm r}<0$, respectively. From the second to the fourth row, solid and dashed curves represent $\gamma>0$ and $\gamma<0$, respectively. }
    \label{fig:plume_case1}
\end{figure}

We run a one-dimensional hybrid simulation for the parameters of Case 1 from Table~\ref{t1}. In this Case, the A/IC is driven unstable during the simulation at frequencies and wavenumbers consistent with those predicted in Section~\ref{ssec:aic.linear} (see Appendix~\ref{sec:disprel_comparison}).
Figure~\ref{fig:energies_case1} (left) shows the evolution of the  total energy variation
\begin{equation}
\Delta W(t)=W_B(t)-W_B(0)+\sum_j\left[W_j(t)-W_j(0)\right],
\end{equation}
the magnetic-energy variation $\Delta W_B=W_B(t)-W_B(0)$ (red), the total kinetic-energy variation $\sum_j\Delta W_j=\sum_j[W_j(t)-W_j(0)]$ (purple), the proton kinetic-energy variation $\Delta W_\text{c}=W_\text{c}(t)-W_\text{c}(0)$ (blue), and the $\alpha$-particle kinetic-energy variation $\Delta W_\alpha=W_\alpha(t)-W_{\alpha}(0)$ (green). All energies are in code units, normalized to $n_\text{c}m_\text{p}v_\text{Ac}^2d_\text{c}^3$. 

The magnetic energy increases at the expense of the system's total kinetic energy, keeping the total energy practically constant (black line). There is a redistribution of particle kinetic energy: $\alpha$-particles transfer energy to the protons while driving the wave unstable. 

In Fig.~\ref{fig:energies_case1} (right), we show the bulk-energy variation $\Delta W_{Uj}=W_{Uj}(t)-W_{Uj}(0)$ (solid), and the thermal-energy variation $\Delta W_{\text{th}j}=W_{\text{th}j}(t)-W_{\text{th}j}(0)$, for each species (blue for protons and green for $\alpha$-particles). Overall, protons gain bulk kinetic and thermal energy as the simulation evolves, while $\alpha$-particles lose energy in a non-monotonic way. In the initial stage of the instability, the bulk kinetic energy of the $\alpha$-particles increases, while their thermal energy decreases, resulting in a net reduction in total kinetic energy. At later stages of the evolution, the bulk kinetic energy decreases while the thermal energy slowly increases, but it remains below the initial value.

\begin{figure}
    \centering
    \includegraphics[width=\linewidth]{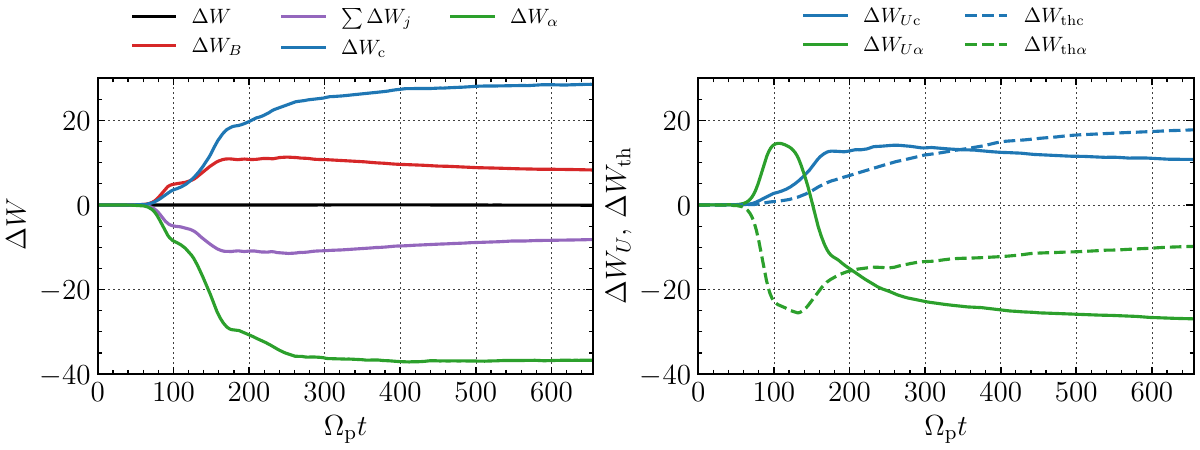}
    \caption{Energy partition from the simulation for Case 1. (left) Evolution of the total-energy variation (black), magnetic-energy variation (red), total kinetic-energy variation (purple), proton kinetic-energy variation (blue), and $\alpha$-particle kinetic-energy variation (green). (right) Changes in bulk ($\Delta W_{\text{U}j}$) and thermal energy ($\Delta W_{\text{th}j}$) for protons and $\alpha$-particles.}
    \label{fig:energies_case1}
\end{figure}

\begin{figure}
    \centering
    \includegraphics[width=\linewidth]{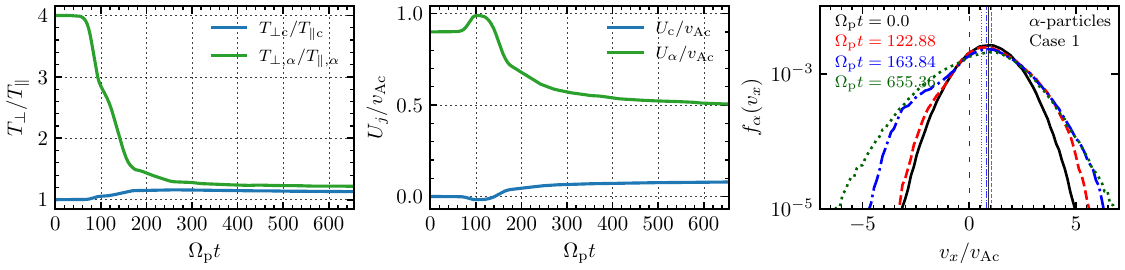}
    \caption{Evolution of the moments for each species: protons (blue) and $\alpha$-particles (green). The left panel shows the temperature anisotropy $T_{\perp j}/T_{\parallel j}$, and the middle panel the parallel drift speed $U_{j}/v_{Ac}$. The right panel shows parallel cuts of the $\alpha$-particle VDF as $f_\alpha(v_{\parallel},v_{\perp}=0)$ at four different times. We also superimpose vertical lines indicating the mean velocities at each time.}
    \label{fig:hydro_case1}
\end{figure}

The growth of magnetic energy is accompanied by a relaxation of the temperature anisotropy of the $\alpha$-particles, as shown in the left panel of Fig.~\ref{fig:hydro_case1}, and by a decrease in their overall kinetic energy $W_{\alpha}$ (left panel of Fig.~\ref{fig:energies_case1}). At the same time, the protons, initially isotropic and stationary in our reference frame, undergo perpendicular heating and net acceleration along the magnetic field, increasing the proton temperature anisotropy and the bulk and thermal energies $W_{\mathrm Uc}$ and $W_{\mathrm{th}c}$. 

The drift-velocity evolution for the $\alpha$-particles shows the same initial increase as in Fig.~\ref{fig:energies_case1} (left panel). We also show field-parallel cuts of the $\alpha$-particle VDF at four different times to demonstrate the evolution of the $\alpha$-particle drift. We superimpose the mean velocity from these cuts as vertical lines. We observe that at $\Omega_\text{p}t=122.88$, the red line shows a more asymmetric distribution, with a tail that is skewed towards the right. This skewness increases the overall bulk velocity of the $\alpha$-particles at that time, showing that this increase in drift speed does not correspond to uniform acceleration of the $\alpha$-particle population. Moments later, at $\Omega_\text{p}t=163.84$ (blue curve), the VDF becomes more symmetric, with a noticeable deformation developing on the left-hand side corresponding to a net reduction in the mean velocity. By the end of the simulation (green curve), the mean velocity has decreased further, resulting in a much more symmetric VDF. The simultaneous decrease in $\alpha$-particle kinetic energy $W_\alpha$ and increase in proton kinetic energy $W_\text{c}$ provide direct evidence that the instability mediates the transfer of energy and momentum from the $\alpha$-particle population to the protons through resonant wave--particle interactions, while the wave grows (i.e., $\delta W_{EB}>0$).

\begin{figure}
    \centering
    \includegraphics[width=0.48\linewidth]{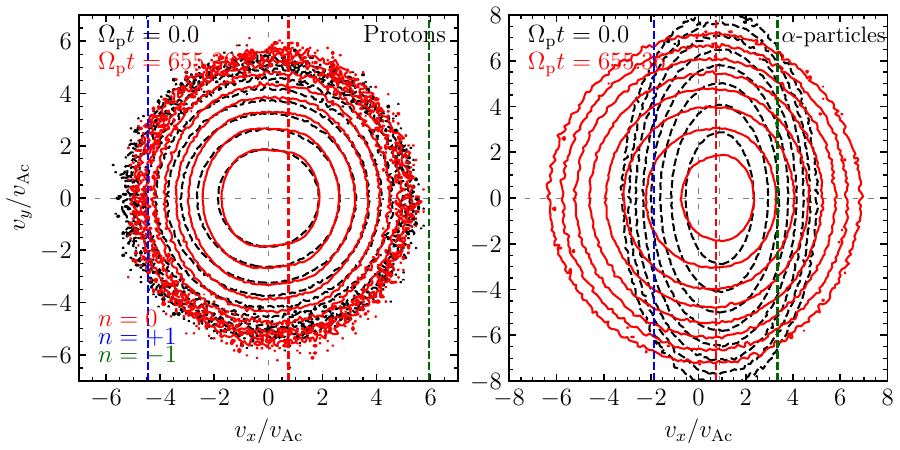}
    \includegraphics[width=0.48\linewidth]{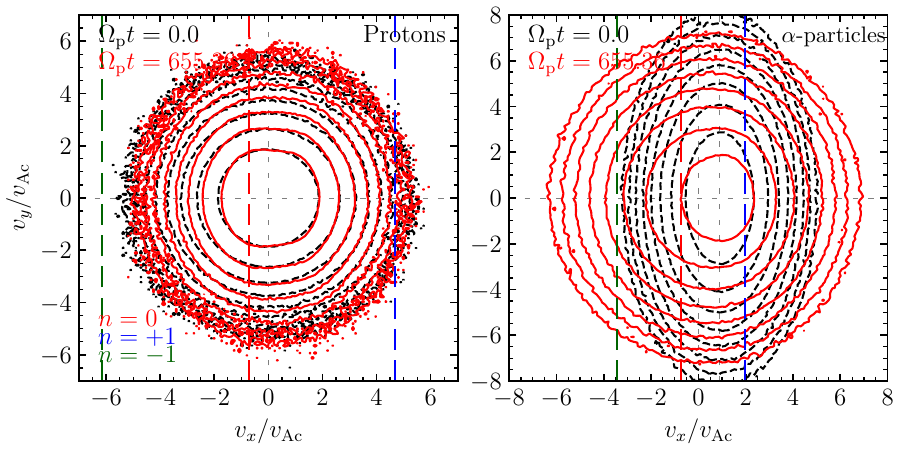}
    \caption{Contours of the proton (left panels) and  $\alpha$-particle (right panels) VDFs in the $(v_x,v_y)$ plane at the initial time ($\Omega_\text{p} t=0$), black dashed contours) and final simulation time ($\Omega_\text{p} t=655.36$), red contours) for Case~1. The vertical dashed lines indicate the resonant velocities associated with the $n=0$ (Landau), $n=+1$, and $n=-1$ resonances for the two unstable modes: A/IC A (left panels) and A/IC B (right panels). }
    \label{fig:fdist_case1}
\end{figure}

Figure~\ref{fig:fdist_case1} shows the evolution of the proton and $\alpha$-particle VDFs between the initial and final stages of the simulation, together with the resonant velocities corresponding to the $n=0$ (Landau; red), $n=+1$ (blue), and $n=-1$ (green) resonances according to Eq.~(\ref{eq:rescond}) for the two unstable modes shown in Fig.~\ref{fig:plume_case1}. The $\alpha$-particle VDF, initially strongly anisotropic, undergoes substantial relaxation during the instability evolution, becoming broader in the parallel direction and more isotropic at later times, consistent with Figure~\ref{fig:hydro_case1}. Simultaneously, the proton VDF develops a slightly enhanced anisotropy with $T_{\perp c}>T_{\parallel c}$, consistent with the proton heating and anisotropy growth observed in the evolution of the moments. The strongest deformation of the VDFs occurs near the $v_{\parallel}$ that fulfills the cyclotron resonance, indicating that the nonlinear evolution is governed primarily by resonant wave--particle interactions and associated quasilinear diffusion in velocity space, as shown in the sketch in Fig.~\ref{fig:cartoon} and supported by the analysis of $\gamma_j^n$ in Fig.~\ref{fig:plume_case1}.

\section{Fast-Magnetosonic/Whistler instability}
\label{sec:FMW}

In this section, we analyze the linear and nonlinear behavior of an unstable plasma in which the resonant instability arises from the relative drift between plasma species. As shown in Table~\ref{tab:params} for Case 2, we include a proton beam so that the system now consists of four species: a proton core $(c)$, proton beam $(b)$, 5\% of drifting $\alpha$-particles $(\alpha)$, and electrons $(e)$. We now study the proton-beam-driven FM/W instability, keeping all species isotropic and introducing a large beam drift (in units of the proton-core Alfv\'en speed). 

In this Case, the $\alpha$-particles, while also drifting, do not contribute to the driving of the instability but may receive energy through wave--particle interactions with the unstable waves. Cases 3 and 4 serve as control cases for Case 2, in which we systematically reduce the $\alpha$-particle population concentration (to 1\% in Case 3 and 0\% in Case 4) to analyze the energy distribution between species when the concentration of the energy-absorbing species decreases. As in the previous section, we begin with linear predictions of the unstable modes and the contributions of each ion population to the growth rates, then present the PIC simulation results and compare them with the linear predictions.

\subsection{Linear Predictions}
\label{ssec:fmw.linear}

\begin{figure}
    \centering
    \includegraphics[width=0.8\linewidth]{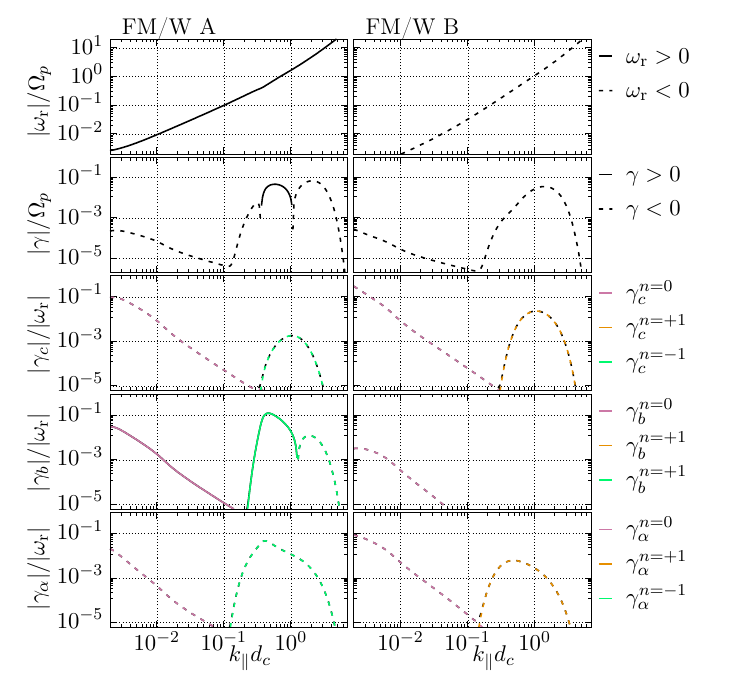}
    \caption{Linear dispersion-relation analysis for Case 2, corresponding to a proton-beam-driven FM/W instability. Format follows Fig.~\ref{fig:plume_case1}, but for the FM/W modes and including an additional, fourth, row for the proton-beam emission and absorption rates.}
    \label{fig:plume_case2}
\end{figure}

For the FM/W instability, we perform a linear dispersion-relation analysis similar to that presented in  Sec.~\ref{ssec:aic.linear}. Fig.~\ref{fig:plume_case2} shows the parallel-propagating FM/W solutions calculated using \texttt{PLUME} for the parameters of Case 2 in Table~\ref{tab:params}. We see in the second row that FM/W B, the backward-propagating solution, is damped (i.e., $\gamma<0$ and thus $\delta W_{EB}<0$) for the entire range of investigated wavevectors. The resonances of the proton-core and $\alpha$-particle populations make a significant contribution to the overall $\gamma$, while the proton beam population does not, leading to enhanced damping near $k_\parallel d_\text{c} \lesssim 1$. For FM/W A, the forward-propagating solution, the proton beam emits sufficient power to overcome the damping by the proton-core and $\alpha$-particle populations to drive a strong instability near $k_\parallel d_\text{c} \sim 0.5$. At these wavenumbers, $\delta W_\text{b}+\delta W_\text{c}+\delta W_\alpha<0$ even though $\delta W_\text{c}>0$ and $\delta W_\alpha>0$.

\begin{figure}
    \centering
    \includegraphics[width=\linewidth]{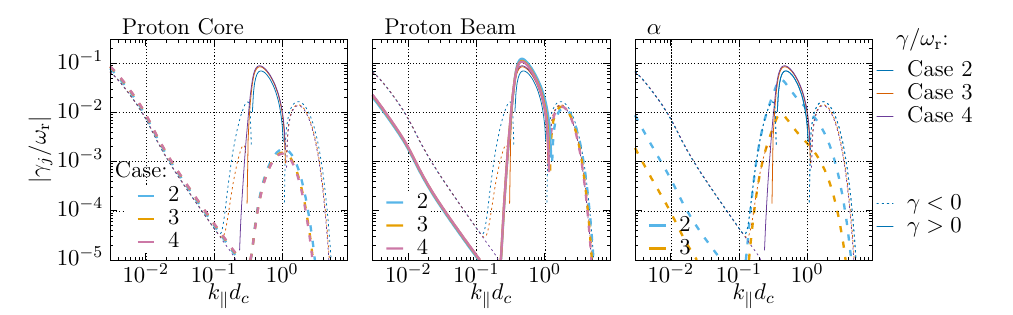}
    \caption{Summary of the growth/damping rate for the proton-beam-driven FM/W instability in Cases 2, 3, and 4. The thin curves show the total $\gamma$, while the thicker curves correspond to the proton core, proton beam, and $\alpha$-particle contributions, respectively. Solid curves denote positive contributions to the wave growth ($\gamma_j>0$), whereas dashed curves indicate contributions to damping ($\gamma_j<0$).}
    \label{fig:plume_summary}
\end{figure}

The relative damping or emission contributions from the three ion populations for FM/W A for Cases 2, 3, and 4 are shown in Fig.~\ref{fig:plume_summary}. In all three Cases, the proton-beam emission is quantitatively similar in wavevector dependence and amplitude. As we reduce the $\alpha$-particle density from Case 2 to Cases 3 and 4, we see a reduction in the magnitude of the $\alpha$-particle contribution $\gamma_\alpha$ and an increase in the mode's total growth rate $\gamma$.

\subsection{Simulation Results}
\label{ssec:fmw.simulation}

We run a nonlinear simulation for the parameters of Cases 2, 3, and 4 in Table~\ref{t1}. In all these Cases, the FM/W instability is driven at frequencies and wavenumbers consistent with those predicted in Section~\ref{ssec:fmw.linear} (see Appendix~\ref{sec:disprel_comparison}). Similar to Case~1 of the A/IC instability, we expect the electromagnetic energy $W_{EB}$ to increase at the expense of the total kinetic energy of the particles $\sum_j W_j$, while the energy of each species $W_j$ may increase or decrease in time due to wave--particle interactions.

\begin{figure}
    \centering
    \includegraphics[width=\linewidth]{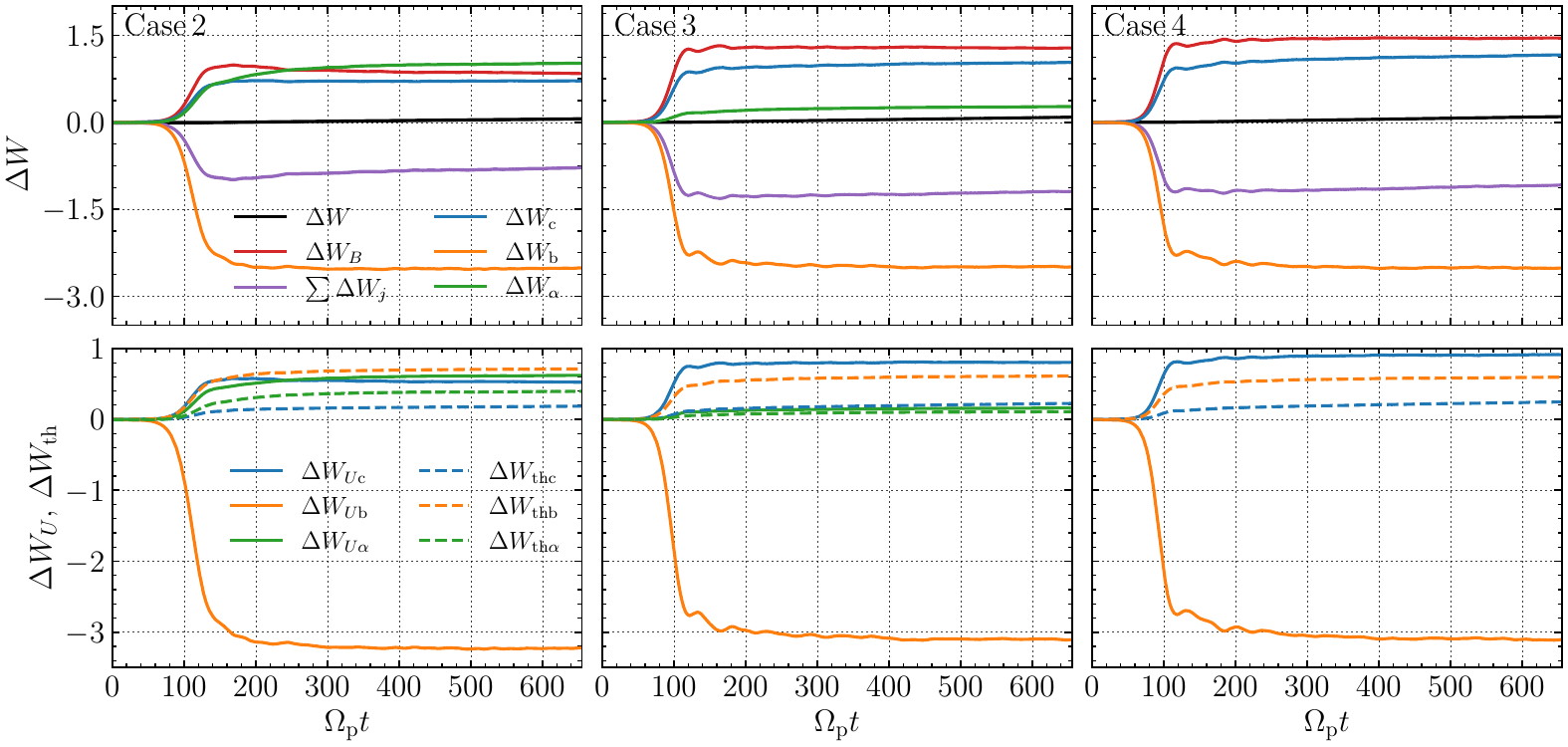}
    \caption{Evolution of the energy partition from the simulation for Cases 2, 3, and 4, in the left, middle, and right panels, respectively. }
    \label{fig:energies_cases234}
\end{figure}

We summarize the energy evolution obtained from the simulations for the FM/W instability in Fig.~\ref{fig:energies_cases234}. Cases 2, 3, and 4 are shown in the left, middle, and right columns. The top panels show the evolution of the variations of the total (black), magnetic (red), kinetic (purple), proton kinetic (blue), beam kinetic (orange), and $\alpha$-particle kinetic energies (green), while the lower panels show the separation of kinetic energy into bulk kinetic and thermal energies (with $\alpha$-particles shown only for Cases 2 and 3). 

As the free energy for this instability is provided by the proton beam, we observe a reduction in the beam kinetic energy $W_\text{b}$, along with increases in the magnetic energy $W_B$ and the core and $\alpha$-particle kinetic energies $W_\text{c}$ and $W_\alpha$. This increase in magnetic energy is translated into a decrease in the total kinetic energy $\sum_jW_j$. 

The comparison of the time evolution of the magnetic energy (red line in the first row) shows that, as the $\alpha$-particle density decreases, the instability becomes more effective, redirecting more of the initial free energy to the fields. Similarly, the higher the $\alpha$-particle density, the larger the fraction of the kinetic energy transferred to them, primarily from the beam. The beam primarily provides energy as bulk kinetic energy $W_{\mathrm Ub}$, as shown in the lower panels, while the thermal energy and bulk kinetic energy of the other species increase.

\begin{figure*}[ht!]
\includegraphics[width=\textwidth]{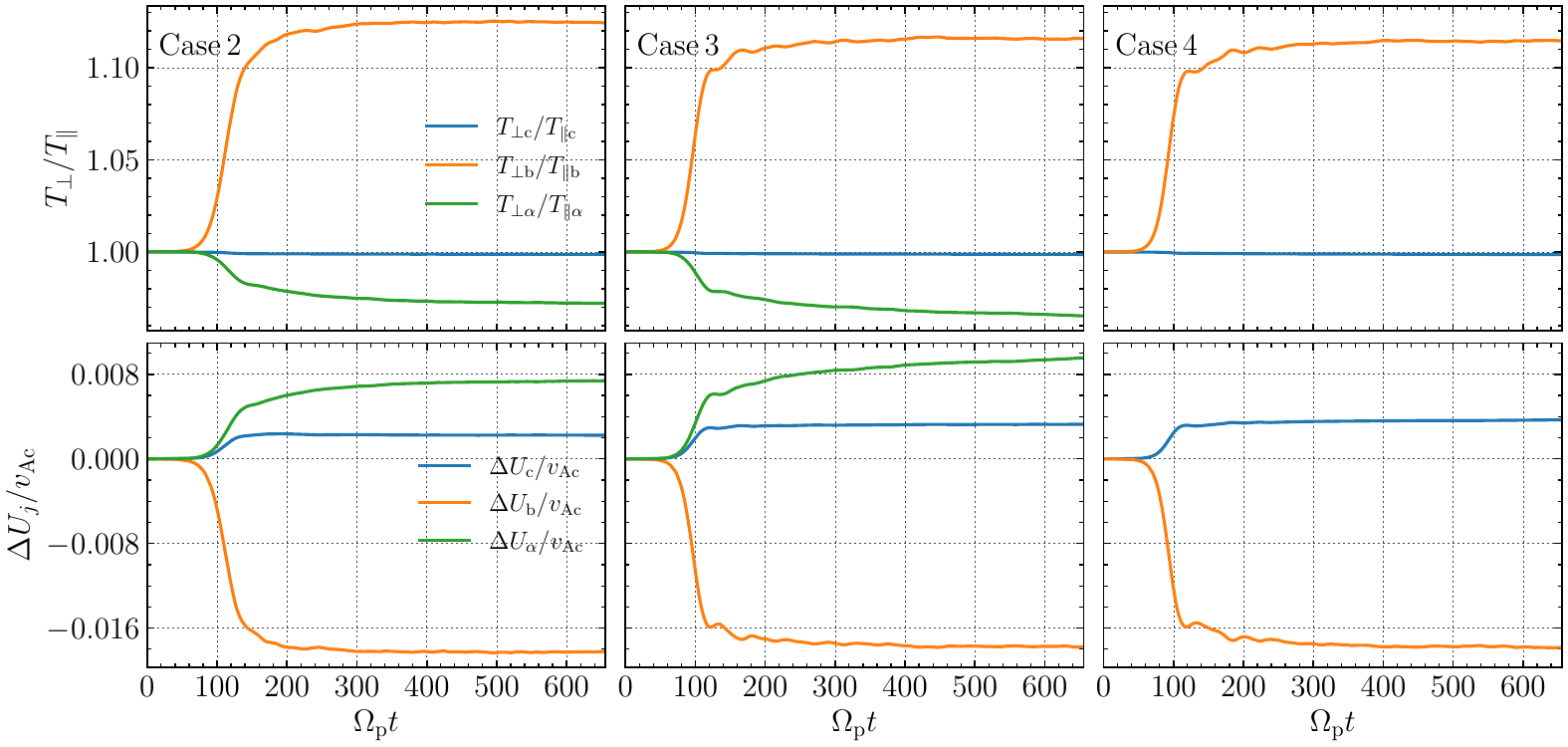}
\caption{Evolution of the anisotropy and drift variation from the simulation for Cases 2, 3, and 4.
\label{fig:hydro_cases234}}
\end{figure*}

Figure~\ref{fig:hydro_cases234} shows the evolution in the temperature anisotropies of the ion species (top panels) and the change in their drift speeds (lower panels). In all three Cases, the proton beam develops a modest perpendicular anisotropy while losing drift speed as the instability extracts energy from the beam's bulk kinetic energy. The proton core remains nearly isotropic in all Cases, although it develops a small positive drift, indicating that a small part of the beam momentum is transferred to the core protons. The $\alpha$-particles develop a slight anisotropy with $T_{\perp \alpha}<T_{\parallel \alpha}$, while the drift speed increases. These changes depend on the $\alpha$-particle concentration. These results indicate that the $\alpha$-particle population modifies the instability's nonlinear saturation by providing an additional resonant channel for the redistribution of energy.

\section{Discussion and Conclusions}
\label{sec:discussion}

In this work, we analyze the generation and relaxation of kinetic micro-instabilities in collisionless multi-species plasmas, and describe inter-species coupling through wave--particle interactions. We perform a linear Vlasov--Maxwell mode analysis and PIC simulations of the A/IC and FM/W instabilities in plasmas composed of protons and $\alpha$-particles with plasma parameters similar to those observed in the solar wind.

Our linear and nonlinear analyses reveal a complex interplay between the particle species and the unstable waves in resonant wave--particle interactions. Depending on the VDFs of the involved particle species $j$ at the relevant resonance velocities, the diffusive flux of particles in velocity space modifies the shape of the VDFs according to quasilinear theory. As long as the resonant wave--particle interactions lead to a net $\sum_j\delta W_j<0$, the electromagnetic field experiences $\delta W_{EB}>0$, corresponding to the unstable growth of the waves. In plasmas with multiple species and even under overall unstable conditions, some species may have $\delta W_j>0$ while others have $\delta W_j<0$. In these cases, the wave--particle interactions effectively transfer energy between the involved particle species while still retaining sufficient energy for the growth of the unstable modes.

In our example Cases, bi-Maxwellian initial VDFs $f_j$ create unstable configurations through a combination of relative drifts and temperature anisotropies. We illustrate with examples in which $\alpha$-particles or a proton beam provide the main free energy to drive instabilities. Our simulations confirm the energetic arguments proposed in Section~\ref{sec:energetics}, in which the overall energy budget allows for transfer between particle species during the growth of an instability. While our work presents four selected example Cases to showcase inter-species coupling, similar couplings are likely relevant to many other configurations, including minor ion species \citep[e.g.,][]{Zhang2025}. Indeed,~\citet{Moya2014} conclude based on a similar analysis including $\alpha$-particles and O$^{+6}$ ions that the presence of anisotropic and drifting ion populations triggers an energy cascade from low to high frequencies, peaking near their corresponding resonance frequencies, with subsequent heating of ions as the charge-to-mass ratio increases. 

The evolution of all $f_j$ ultimately leads to a marginally stable wave--particle equilibrium state for which the overall growth/damping rate vanishes. In this quasilinear saturation state, the $f_j$ of all involved particle species have -- locally in velocity space as a result of quasilinear diffusion -- adjusted to a configuration in which continuing wave--particle interactions keep $\delta W_{j}=0$ for all $j$ and thus $\delta W_{EB}=0$ \citep{Villarroel2026}. These wave--particle equilibria of all species with significant $f_j$ distortion at the relevant resonance speeds are an observable prediction of our study \citep{Walters2023,Ran2026}. The analysis of observed $f_j$ with the help of the ALPS code \citep{Verscharen2018,Klein2025}, which evaluates the linear Vlasov--Maxwell dispersion relation without the assumption of bi-Maxwellian VDFs, has the potential to reveal the presence of such wave--particle equilibria.

Our analysis makes limiting assumptions on the plasma configurations under consideration. For instance, we assume that the background plasma is homogeneous and in a steady state. The solar wind, however, is an expanding and turbulent medium \citep{Matteini2012,Bruno2013}, suggesting that the steady-state and homogeneity assumptions require careful consideration and justification based on arguments of scale separation \citep{Alexandrova2013,Hellinger2015,Verscharen2016,Hellinger2019,Zhu2023}. Observations of the solar wind reveal, however, that predictions for instabilities based on linear Vlasov--Maxwell theory are often appropriate even in such a nonlinear medium \citep{Klein2018,Opie2022,Opie2024}. Moreover, our analysis so far accounts for electrons either as an isotropic Maxwellian (in our linear calculations) or as a charge-neutralizing, massless fluid (in our hybrid simulations). In reality, though, electrons also exhibit a wide variety of non-equilibrium features \citep{Feldman1975,Pilipp1987,Halekas2020} that can drive kinetic micro-instabilities \citep{Verscharen2022}. It would be worthwhile to extend our study to examine the effects of non-Maxwellian electron background VDFs in both linear theory and full particle-in-cell simulations.

\begin{acknowledgments}
This research was supported by the International Space Science
Institute (ISSI) in Bern, through ISSI International Team project
``Excitation and dissipation of kinetic‐scale fluctuations in space plasmas" (ISSI Team project \#24‐612). We acknowledge the support of ANID Chile through FONDECyT grant No.\ 1251712 (R.A.L.) and 1240281 (P.S.M.). K.G.K. was partially supported by NASA grant 80NSSC24K0171.  A.T. was partly supported by U.S. National Science Foundation PHY-2010189 and the U.S. Department of Energy (DOE) Fusion Energy Sciences Postdoctoral Research Program, administered by the Oak Ridge Institute for Science and Education and Oak Ridge Associated Universities under DOE contract DE-SC0014664. D.V. is supported by STFC Consolidated Grant ST/W001004/1.
High-performance computing resources were provided by Anvil at Purdue University's Rosen Center for Advanced Computing via allocation PHY250393 from the U.S. National Science Foundation (NSF) Advanced Cyberinfrastructure Coordination Ecosystem: Services \& Support (ACCESS) program. Powered@NLHPC: This research was partially supported by the supercomputing infrastructure of the NLHPC (CCSS210001).
The ALPS project received support from UCL’s Advanced Research Computing Centre through the Open Source Software Sustainability Funding scheme.
\end{acknowledgments}


\appendix

\section{Energy conservation}
\label{sec:appendix_conservation}

\begin{figure}
    \centering
    \includegraphics[width=0.7\linewidth]{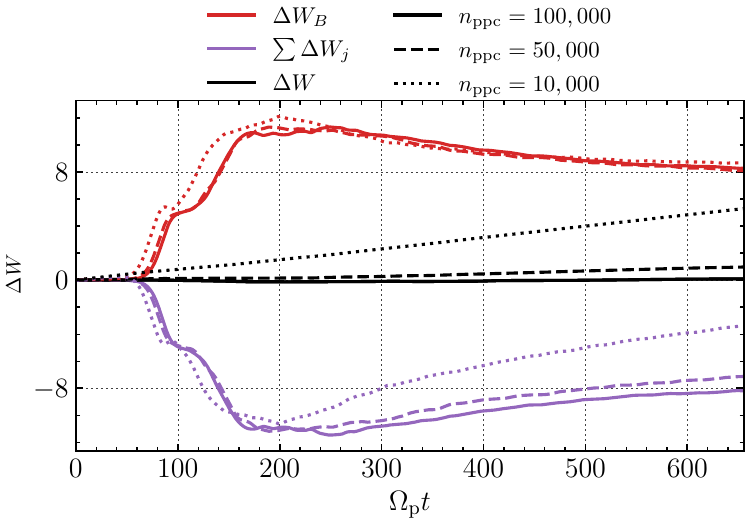}
    \caption{Energy conservation test for Case 1. Evolution of the variations of total energy (black), magnetic energy (red), and total kinetic energy (purple) for three different numbers of particles per cell per species: Solid lines for $n_\text{ppc}=100,000$, dashed lines for $n_\text{ppc}=50,000$, and dotted lines for $n_\text{ppc}=10,000$. }
    \label{fig:energy_conservation}
\end{figure}

Since we are interested in the energy partition among plasma species, we must ensure that the total energy is conserved in our simulations. We take Case 1 from Table~\ref{t1} and run three simulations with varying numbers of particles per cell per species. We use $n_\text{ppc}=10,000$ (dotted lines), $50,000$ (dashed), and $100,000$ (solid). Figure~\ref{fig:energy_conservation} shows the evolution of the variations in the total energy (black), magnetic energy (red), and total kinetic energy (purple) depending on $n_{\text{ppc}}$. 

Increasing $n_{\text{ppc}}$ ensures better energy conservation. The main differences are observed in the total kinetic energy, while the magnetic energy is almost independent of $n_{\text{ppc}}$. The overall trend in energy evolution is maintained, even when energy conservation is poor.

\section{Nonlinear Spectra and Dispersion Relation}
\label{sec:disprel_comparison}

\begin{figure}
    \centering
    \includegraphics[width=0.9\linewidth]{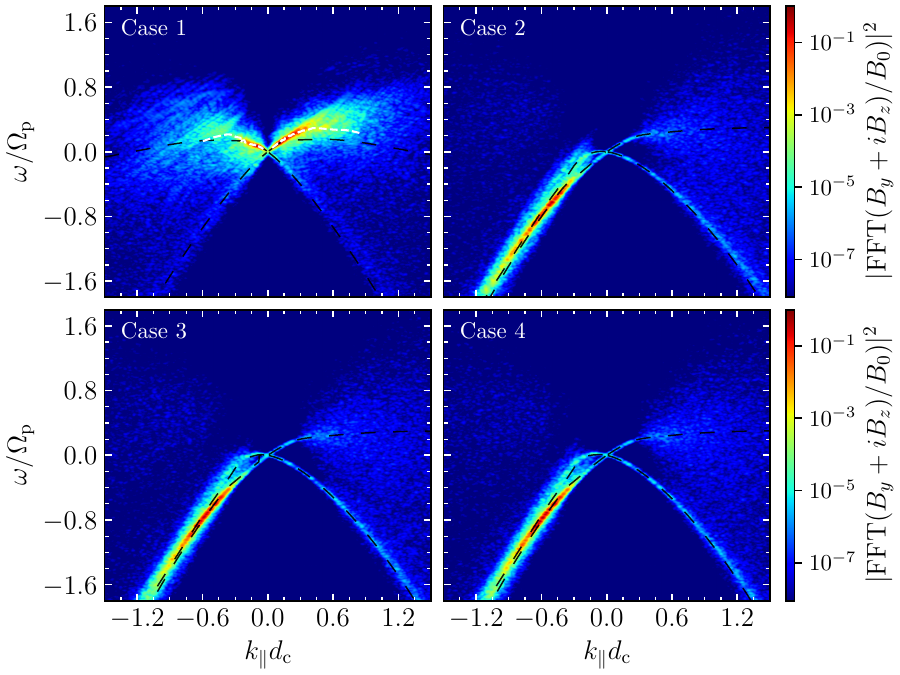}
    \caption{Dispersion relation obtained from $\text{FFT}(B_y+iB_z)/B_0$ applied to the simulation data for all four Cases. The first and second quadrants correspond to left-hand circularly polarized modes in forward and backward propagation, respectively, whereas the third and fourth quadrants correspond to right-hand circularly polarized modes in forward and backward propagation, respectively.}
    \label{fig:spectras}
\end{figure}

In Fig.~\ref{fig:spectras}, we compare the dispersion relation from the simulation with the predictions by linear theory. We perform the Fast Fourier Transform (FFT) of the complete temporal series for the magnetic field fluctuations of our simulation data as $\text{FFT}(B_y+iB_z)/B_0$. The first and second quadrants correspond to left-hand circularly polarized modes in forward and backward propagation, respectively, whereas the third and fourth quadrants correspond to right-hand circularly polarized modes in forward and backward propagation, respectively. We normalize Case 1 to the maximum of its power spectrum. For Cases 2, 3, and 4, we normalize all spectra to their common maximum, which corresponds to the maximum of Case 4. This ordering is consistent with the linear theory prediction in Fig.~\ref{fig:plume_summary}. 

Over the FFT results from our simulations, we superimpose the linear dispersion-relation solutions as black dashed lines on each spectrum, using the initial-condition parameters for each Case. The spectra agree well with the predicted linear normal modes of the plasma, as well as the unstable modes that develop throughout the simulation. For Case 1, the unstable modes lie in the first and second quadrants, consistent with the A/IC instability. Cases 2, 3, and 4 show the unstable branch in the third quadrant, consistent with the FM/W instability. 

The linear prediction for Case 1 differs from the simulation spectrum slightly in that the real frequency is below the simulation spectrum for the A/IC branch in the first quadrant. This difference arises because the linear-theory prediction is a snapshot based on the initial condition, whereas the simulation spectra capture the entire simulation time window, and, as seen in Fig.~\ref{fig:fdist_case1}, the VDFs evolve significantly during the simulation. 

To corroborate this claim, we use the ALPS solver~\citep{Verscharen2018} to compute the linear dispersion relation using a Chebyshev polynomial representation of the actual proton and $\alpha-$particle VDFs obtained from the simulation at a representative time of $\Omega_\text{p}t=409.60$; details of the Chebyshev representation are given by \cite{Klein2025}. By this simulation time, the instability has mostly saturated, and the system is close to a steady wave--particle equilibrium state, as we show in Figure~\eqref{fig:energies_case1}.  We add this solution as a white dashed line to the Case 1 spectrum, which we plot up to the wavenumber where $|\gamma|/|\omega_{\mathrm{r}}|=1$ to exclude strongly damped modes from our visualization. The observed late-time simulation spectrum for Case 1 is consistent with the dispersion relation obtained from ALPS. The agreement between the two is most notable at low wavenumbers until the wavenumber at which the wave power begins to decrease again. Although the dispersion relations deviate from the simulation spectrum at larger wavenumbers, the difference in the wavenumber ranges that satisfy the conditions $|\gamma|/|\omega_{\mathrm{r}}|<1$ for forward- and backward-propagating A/IC waves can also explain the reflectional asymmetry of the observed spectra. 

\bibliography{biblio}{}

@article{Le2023,
   author = {Ari Le and Adam Stanier and Lin Yin and Blake Wetherton and Brett Keenan and Brian Albright},
   doi = {10.1063/5.0146529},
   issn = {1070-664X},
   issue = {6},
   journal = {Physics of Plasmas},
   month = {6},
   pages = {063902},
   publisher = {AIP Publishing LLC},
   title = {Hybrid-VPIC: An open-source kinetic/fluid hybrid particle-in-cell code},
   volume = {30},
   url = {https://pubs.aip.org/pop/article/30/6/063902/2893917/Hybrid-VPIC-An-open-source-kinetic-fluid-hybrid},
   year = {2023}
}

@article{Le2021,
   author = {Ari Le and Dan Winske and Adam Stanier and William Daughton and Misa Cowee and Blake Wetherton and Fan Guo},
   doi = {10.1029/2021JA029125},
   issn = {2169-9380},
   issue = {9},
   journal = {Journal of Geophysical Research: Space Physics},
   month = {9},
   pages = {1-15},
   title = {Astrophysical Explosions Revisited: Collisionless Coupling of Debris to Magnetized Plasma},
   volume = {126},
   url = {https://agupubs.onlinelibrary.wiley.com/doi/10.1029/2021JA029125},
   year = {2021}
}

@article{Bowers2008,
   author = {K. J. Bowers and B. J. Albright and L. Yin and B. Bergen and T. J. T. Kwan},
   doi = {10.1063/1.2840133},
   issn = {1070-664X},
   issue = {5},
   journal = {Physics of Plasmas},
   month = {5},
   pages = {055703},
   title = {Ultrahigh performance three-dimensional electromagnetic relativistic kinetic plasma simulation},
   volume = {15},
   url = {https://pubs.aip.org/pop/article/15/5/055703/1015800/Ultrahigh-performance-three-dimensional},
   year = {2008}
}

@misc{PLUME_2025,
  author       = {{Klein}, K. G. and
                  {Howes}, G. G.},
  title        = {kgklein/PLUME: Zenodo release},
  year         = 2025,
  publisher    = {Zenodo},
  version      = {v1.0.1},
  doi          = {10.5281/zenodo.15215514},
  url          = {https://doi.org/10.5281/zenodo.15215514}
}

@article{Klein_2025,
doi = {10.3847/2515-5172/add1c2},
url = {https://doi.org/10.3847/2515-5172/add1c2},
year = {2025},
month = {apr},
publisher = {The American Astronomical Society},
volume = {9},
number = {4},
pages = {102},
author = {Klein, Kristopher G. and Howes, Gregory G. and Brown, Collin R.},
title = {PLUME: Plasma in a Linear Uniform Magnetized Environment},
journal = {Research Notes of the AAS},
}

@ARTICLE{Verscharen2019,
       author = {{Verscharen}, Daniel and {Klein}, Kristopher G. and {Maruca}, Bennett A.},
        title = "{The multi-scale nature of the solar wind}",
      journal = {Living Reviews in Solar Physics},
         year = 2019,
        month = dec,
       volume = {16},
       number = {1},
          eid = {5},
        pages = {5},
          doi = {10.1007/s41116-019-0021-0},
archivePrefix = {arXiv},
       eprint = {1902.03448},
 primaryClass = {physics.space-ph},
       adsurl = {https://ui.adsabs.harvard.edu/abs/2019LRSP...16....5V}
}

@ARTICLE{Marsch2006,
       author = {{Marsch}, Eckart},
        title = "{Kinetic Physics of the Solar Corona and Solar Wind}",
      journal = {Living Reviews in Solar Physics},
         year = 2006,
        month = dec,
       volume = {3},
       number = {1},
          eid = {1},
        pages = {1},
          doi = {10.12942/lrsp-2006-1},
       adsurl = {https://ui.adsabs.harvard.edu/abs/2006LRSP....3....1M}
}

@ARTICLE{Marsch1982,
       author = {{Marsch}, E. and {Schwenn}, R. and {Rosenbauer}, H. and {Muehlhaeuser}, K.-H. and {Pilipp}, W. and {Neubauer}, F.~M.},
        title = "{Solar wind protons: Three-dimensional velocity distributions and derived plasma parameters measured between 0.3 and 1 AU}",
      journal = {\jgr},
         year = 1982,
        month = jan,
       volume = {87},
       number = {A1},
        pages = {52-72},
          doi = {10.1029/JA087iA01p00052},
       adsurl = {https://ui.adsabs.harvard.edu/abs/1982JGR....87...52M}
}

@ARTICLE{Hundhausen1967,
       author = {{Hundhausen}, A.~J. and {Bame}, S.~J. and {Ness}, N.~F.},
        title = "{Solar Wind Thermal Anisotropies: Vela 3 and IMP 3}",
      journal = {\jgr},
         year = 1967,
        month = nov,
       volume = {72},
        pages = {5265},
          doi = {10.1029/JZ072i021p05265},
       adsurl = {https://ui.adsabs.harvard.edu/abs/1967JGR....72.5265H}
}

@ARTICLE{Goodrich1976,
       author = {{Goodrich}, C.~C. and {Lazarus}, A.~J.},
        title = "{Suprathermal protons in the interplanetary solar wind}",
      journal = {\jgr},
         year = 1976,
        month = jun,
       volume = {81},
       number = {16},
        pages = {2750},
          doi = {10.1029/JA081i016p02750},
       adsurl = {https://ui.adsabs.harvard.edu/abs/1976JGR....81.2750G}
}

@article{Espinoza2018,
author = {Espinoza, C. M. and Stepanova, M. and Moya, P. S. and Antonova, E. E. and Valdivia, J. A.},
title = {Ion and Electron κ Distribution Functions Along the Plasma Sheet},
journal = {Geophysical Research Letters},
volume = {45},
number = {13},
pages = {6362-6370},
doi = {https://doi.org/10.1029/2018GL078631},
url = {https://agupubs.onlinelibrary.wiley.com/doi/abs/10.1029/2018GL078631},
eprint = {https://agupubs.onlinelibrary.wiley.com/doi/pdf/10.1029/2018GL078631},
year = {2018}
}

@article{Christon1988,
author = {Christon, S. P. and Mitchell, D. G. and Williams, D. J. and Frank, L. A. and Huang, C. Y. and Eastman, T. E.},
title = {Energy spectra of plasma sheet ions and electrons from ∼50 eV/e to ∼1 MeV during plasma temperature transitions},
journal = {Journal of Geophysical Research: Space Physics},
volume = {93},
number = {A4},
pages = {2562-2572},
doi = {https://doi.org/10.1029/JA093iA04p02562},
url = {https://agupubs.onlinelibrary.wiley.com/doi/abs/10.1029/JA093iA04p02562},
eprint = {https://agupubs.onlinelibrary.wiley.com/doi/pdf/10.1029/JA093iA04p02562},
year = {1988}
}

@article{Eyelade2025,
  title = {Characterizing solar wind electrons with the core-strahlo model: WIND-SWE-VEIS observations},
  volume = {702},
  ISSN = {1432-0746},
  url = {http://dx.doi.org/10.1051/0004-6361/202555368},
  DOI = {10.1051/0004-6361/202555368},
  journal = {Astronomy \& Astrophysics},
  publisher = {EDP Sciences},
  author = {Eyelade,  Adetayo V. and Zenteno-Quinteros,  Bea and Moya,  Pablo S. and Silva,  Javier I. and Urra,  Benjamin A. and Lazar,  Marian and Viñas,  Adolfo F.},
  year = {2025},
  month = Oct,
  pages = {A198}
}

@ARTICLE{Asbridge1976,
       author = {{Asbridge}, J.~R. and {Bame}, S.~J. and {Feldman}, W.~C. and {Montgomery}, M.~D.},
        title = "{Helium and hydrogen velocity differences in the solar wind}",
      journal = {\jgr},
         year = 1976,
        month = jun,
       volume = {81},
       number = {16},
        pages = {2719},
          doi = {10.1029/JA081i016p02719},
       adsurl = {https://ui.adsabs.harvard.edu/abs/1976JGR....81.2719A}
}

@ARTICLE{Hellinger2006,
       author = {{Hellinger}, Petr and {Tr{\'a}vn{\'\i}{\v{c}}ek}, Pavel and {Kasper}, Justin C. and {Lazarus}, Alan J.},
        title = "{Solar wind proton temperature anisotropy: Linear theory and WIND/SWE observations}",
      journal = {\grl},
         year = 2006,
        month = may,
       volume = {33},
       number = {9},
          eid = {L09101},
        pages = {L09101},
          doi = {10.1029/2006GL025925},
       adsurl = {https://ui.adsabs.harvard.edu/abs/2006GeoRL..33.9101H}
}

@ARTICLE{Gary1991,
       author = {{Gary}, S. Peter},
        title = "{Electromagnetic Ion / Ion Instabilities and Their Consequences in Space Plasmas - a Review}",
      journal = {\ssr},
         year = 1991,
        month = may,
       volume = {56},
       number = {3-4},
        pages = {373-415},
          doi = {10.1007/BF00196632},
       adsurl = {https://ui.adsabs.harvard.edu/abs/1991SSRv...56..373G}
}

@ARTICLE{Matteini2007,
       author = {{Matteini}, Lorenzo and {Landi}, Simone and {Hellinger}, Petr and {Pantellini}, Filippo and {Maksimovic}, Milan and {Velli}, Marco and {Goldstein}, Bruce E. and {Marsch}, Eckart},
        title = "{Evolution of the solar wind proton temperature anisotropy from 0.3 to 2.5 AU}",
      journal = {\grl},
         year = 2007,
        month = oct,
       volume = {34},
       number = {20},
          eid = {L20105},
        pages = {L20105},
          doi = {10.1029/2007GL030920},
       adsurl = {https://ui.adsabs.harvard.edu/abs/2007GeoRL..3420105M}
}

@ARTICLE{Maruca2012,
       author = {{Maruca}, Bennett A. and {Kasper}, Justin C. and {Gary}, S. Peter},
        title = "{Instability-driven Limits on Helium Temperature Anisotropy in the Solar Wind: Observations and Linear Vlasov Analysis}",
      journal = {\apj},
         year = 2012,
        month = apr,
       volume = {748},
       number = {2},
          eid = {137},
        pages = {137},
          doi = {10.1088/0004-637X/748/2/137},
       adsurl = {https://ui.adsabs.harvard.edu/abs/2012ApJ...748..137M}
}

@ARTICLE{Verniero2020,
       author = {{Verniero}, J.~L. and {Larson}, D.~E. and {Livi}, R. and {Rahmati}, A. and {McManus}, M.~D. and {Pyakurel}, P. Sharma and {Klein}, K.~G. and {Bowen}, T.~A. and {Bonnell}, J.~W. and {Alterman}, B.~L. and et al.},
        title = "{Parker Solar Probe Observations of Proton Beams Simultaneous with Ion-scale Waves}",
      journal = {\apjs},
         year = 2020,
        month = may,
       volume = {248},
       number = {1},
          eid = {5},
        pages = {5},
          doi = {10.3847/1538-4365/ab86af},
archivePrefix = {arXiv},
       eprint = {2004.03009},
 primaryClass = {physics.space-ph},
       adsurl = {https://ui.adsabs.harvard.edu/abs/2020ApJS..248....5V}
}

@BOOK{Gary1993,
       author = {{Gary}, S. Peter},
        title = "{Theory of Space Plasma Microinstabilities}",
         year = 1993,
       adsurl = {https://ui.adsabs.harvard.edu/abs/1993tspm.book.....G}
}

@ARTICLE{Kennel1966,
       author = {{Kennel}, C.~F. and {Engelmann}, F.},
        title = "{Velocity Space Diffusion from Weak Plasma Turbulence in a Magnetic Field}",
      journal = {Physics of Fluids},
         year = 1966,
        month = dec,
       volume = {9},
       number = {12},
        pages = {2377-2388},
          doi = {10.1063/1.1761629},
       adsurl = {https://ui.adsabs.harvard.edu/abs/1966PhFl....9.2377K}
}

@ARTICLE{Kennel1967,
       author = {{Kennel}, C.~F. and {Wong}, H.~V.},
        title = "{Resonant particle instabilities in a uniform magnetic field}",
      journal = {Journal of Plasma Physics},
         year = 1967,
        month = feb,
       volume = {1},
       number = {1},
        pages = {75-80},
          doi = {10.1017/S002237780000310X},
       adsurl = {https://ui.adsabs.harvard.edu/abs/1967JPlPh...1...75K}
}

@ARTICLE{Daughton1998,
       author = {{Daughton}, William and {Gary}, S. Peter},
        title = "{Electromagnetic proton/proton instabilities in the solar wind}",
      journal = {\jgr},
         year = 1998,
        month = sep,
       volume = {103},
       number = {A9},
        pages = {20613-20620},
          doi = {10.1029/98JA01385},
       adsurl = {https://ui.adsabs.harvard.edu/abs/1998JGR...10320613D}
}

@ARTICLE{Marsch1982b,
       author = {{Marsch}, E. and {Rosenbauer}, H. and {Schwenn}, R. and {Muehlhaeuser}, K.-H. and {Neubauer}, F.~M.},
        title = "{Solar wind helium ions: obsevations of the Helios solar probes between 0.3 and 1 AU}",
      journal = {\jgr},
         year = 1982,
        month = jan,
       volume = {87},
       number = {A1},
        pages = {35-51},
          doi = {10.1029/JA087iA01p00035},
       adsurl = {https://ui.adsabs.harvard.edu/abs/1982JGR....87...35M}
}

@ARTICLE{Goldstein2000,
       author = {{Goldstein}, Bruce E. and {Neugebauer}, Marcia and {Zhang}, L. Dennis and {Gary}, S. Peter},
        title = "{Observed constraint on proton-proton relative velocities in the solar wind}",
      journal = {\grl},
         year = 2000,
        month = jan,
       volume = {27},
       number = {1},
        pages = {53-56},
          doi = {10.1029/1999GL003637},
       adsurl = {https://ui.adsabs.harvard.edu/abs/2000GeoRL..27...53G}
}

@ARTICLE{Verscharen2013,
       author = {{Verscharen}, Daniel and {Bourouaine}, Sofiane and {Chandran}, Benjamin D.~G.},
        title = "{Instabilities Driven by the Drift and Temperature Anisotropy of Alpha Particles in the Solar Wind}",
      journal = {\apj},
         year = 2013,
        month = aug,
       volume = {773},
       number = {2},
          eid = {163},
        pages = {163},
          doi = {10.1088/0004-637X/773/2/163},
archivePrefix = {arXiv},
       eprint = {1307.1823},
 primaryClass = {physics.space-ph},
       adsurl = {https://ui.adsabs.harvard.edu/abs/2013ApJ...773..163V}
}

@ARTICLE{Gary2000,
       author = {{Gary}, S. Peter and {Yin}, Lin and {Winske}, Dan and {Reisenfeld}, Daniel B.},
        title = "{Electromagnetic alpha/proton instabilities in the solar wind}",
      journal = {\grl},
         year = 2000,
        month = may,
       volume = {27},
       number = {9},
        pages = {1355-1358},
          doi = {10.1029/2000GL000019},
       adsurl = {https://ui.adsabs.harvard.edu/abs/2000GeoRL..27.1355G}
}

@ARTICLE{Montgomery1976,
       author = {{Montgomery}, M.~D. and {Gary}, S.~P. and {Feldman}, W.~C. and {Forslund}, D.~W.},
        title = "{Electromagnetic instabilities driven by unequal proton beams in the solar wind}",
      journal = {\jgr},
         year = 1976,
        month = jun,
       volume = {81},
       number = {16},
        pages = {2743},
          doi = {10.1029/JA081i016p02743},
       adsurl = {https://ui.adsabs.harvard.edu/abs/1976JGR....81.2743M}
}

@ARTICLE{Reisenfeld2001,
       author = {{Reisenfeld}, D.~B. and {Gary}, S.~P. and {Gosling}, J.~T. and {Steinberg}, J.~T. and {McComas}, D.~J. and {Goldstein}, B.~E. and {Neugebauer}, M.},
        title = "{Helium energetics in the high-latitude solar wind: Ulysses observations}",
      journal = {\jgr},
         year = 2001,
        month = apr,
       volume = {106},
       number = {A4},
        pages = {5693-5708},
          doi = {10.1029/2000JA000317},
       adsurl = {https://ui.adsabs.harvard.edu/abs/2001JGR...106.5693R}
}

@ARTICLE{Podesta2011,
       author = {{Podesta}, John J. and {Gary}, S. Peter},
        title = "{Effect of Differential Flow of Alpha Particles on Proton Pressure Anisotropy Instabilities in the Solar Wind}",
      journal = {\apj},
         year = 2011,
        month = nov,
       volume = {742},
       number = {1},
          eid = {41},
        pages = {41},
          doi = {10.1088/0004-637X/742/1/41},
       adsurl = {https://ui.adsabs.harvard.edu/abs/2011ApJ...742...41P}
}

@ARTICLE{DeMarco2023,
       author = {{De Marco}, R. and {Bruno}, R. and {Jagarlamudi}, V. Krishna and {D'Amicis}, R. and {Marcucci}, M.~F. and {Fortunato}, V. and {Perrone}, D. and {Telloni}, D. and {Owen}, C.~J. and {Louarn}, P. and et al.},
        title = "{Innovative technique for separating proton core, proton beam, and alpha particles in solar wind 3D velocity distribution functions}",
      journal = {\aap},
         year = 2023,
        month = jan,
       volume = {669},
          eid = {A108},
        pages = {A108},
          doi = {10.1051/0004-6361/202243719},
       adsurl = {https://ui.adsabs.harvard.edu/abs/2023A&A...669A.108D}
}

@ARTICLE{Bruno2024,
       author = {{Bruno}, Roberto and {De Marco}, Rossana and {D'Amicis}, Raffaella and {Perrone}, Denise and {Marcucci}, Maria Federica and {Telloni}, Daniele and {Marino}, Raffaele and {Sorriso-Valvo}, Luca and {Fortunato}, Vito and {Mele}, Gennaro and et al.},
        title = "{Comparative Study of the Kinetic Properties of Proton and Alpha Beams in the Alfv{\'e}nic Wind Observed by SWA-PAS On Board Solar Orbiter}",
      journal = {\apj},
         year = 2024,
        month = jul,
       volume = {969},
       number = {2},
          eid = {106},
        pages = {106},
          doi = {10.3847/1538-4357/ad47b3},
archivePrefix = {arXiv},
       eprint = {2403.10489},
 primaryClass = {astro-ph.SR},
       adsurl = {https://ui.adsabs.harvard.edu/abs/2024ApJ...969..106B}
}

@ARTICLE{Bourouaine2013,
       author = {{Bourouaine}, Sofiane and {Verscharen}, Daniel and {Chandran}, Benjamin D.~G. and {Maruca}, Bennett A. and {Kasper}, Justin C.},
        title = "{Limits on Alpha Particle Temperature Anisotropy and Differential Flow from Kinetic Instabilities: Solar Wind Observations}",
      journal = {\apjl},
         year = 2013,
        month = nov,
       volume = {777},
       number = {1},
          eid = {L3},
        pages = {L3},
          doi = {10.1088/2041-8205/777/1/L3},
archivePrefix = {arXiv},
       eprint = {1309.4010},
 primaryClass = {astro-ph.SR},
       adsurl = {https://ui.adsabs.harvard.edu/abs/2013ApJ...777L...3B}
}

@ARTICLE{Feldman1973,
       author = {{Feldman}, W.~C. and {Asbridge}, J.~R. and {Bame}, S.~J. and {Montgomery}, M.~D.},
        title = "{Double ion streams in the solar wind}",
      journal = {\jgr},
         year = 1973,
        month = jan,
       volume = {78},
       number = {13},
        pages = {2017},
          doi = {10.1029/JA078i013p02017},
       adsurl = {https://ui.adsabs.harvard.edu/abs/1973JGR....78.2017F}
}

@ARTICLE{Tu2002,
       author = {{Tu}, C.-Y. and {Marsch}, E.},
        title = "{Anisotropy regulation and plateau formation through pitch angle diffusion of solar wind protons in resonance with cyclotron waves}",
      journal = {Journal of Geophysical Research (Space Physics)},
         year = 2002,
        month = sep,
       volume = {107},
       number = {A9},
          eid = {1249},
        pages = {1249},
          doi = {10.1029/2001JA000150},
       adsurl = {https://ui.adsabs.harvard.edu/abs/2002JGRA..107.1249T}
}

@ARTICLE{Rowlands1966,
       author = {{Rowlands}, J. and {Shapiro}, V.~D. and {Shevchenko}, V.~I.},
        title = "{Quasilinear Theory of Plasma Cyclotron Instability}",
      journal = {Soviet Journal of Experimental and Theoretical Physics},
         year = 1966,
        month = oct,
       volume = {23},
        pages = {651},
       adsurl = {https://ui.adsabs.harvard.edu/abs/1966JETP...23..651R}
}

@INPROCEEDINGS{Shapiro1966,
       author = {{Shapiro}, V.~D. and {Shevchenko}, V.~I.},
        title = "{Quasilinear theory of instability of a plasma with anisotropic ion velocity distribution}",
    booktitle = {Plasma Physics and Controlled Thermonuclear Fusion},
         year = 1966,
       editor = {{Sinel'Nikov}, K.~D.},
        month = jan,
        pages = {149},
       adsurl = {https://ui.adsabs.harvard.edu/abs/1966ppct.conf..149S}
}

@ARTICLE{Marsch1987,
       author = {{Marsch}, E. and {Livi}, S.},
        title = "{Observational evidence for marginal stability of solar wind ion beams}",
      journal = {\jgr},
         year = 1987,
        month = jul,
       volume = {92},
       number = {A7},
        pages = {7263-7268},
          doi = {10.1029/JA092iA07p07263},
       adsurl = {https://ui.adsabs.harvard.edu/abs/1987JGR....92.7263M}
}

@ARTICLE{Bale2009,
       author = {{Bale}, S.~D. and {Kasper}, J.~C. and {Howes}, G.~G. and {Quataert}, E. and {Salem}, C. and {Sundkvist}, D.},
        title = "{Magnetic Fluctuation Power Near Proton Temperature Anisotropy Instability Thresholds in the Solar Wind}",
      journal = {\prl},
         year = 2009,
        month = nov,
       volume = {103},
       number = {21},
          eid = {211101},
        pages = {211101},
          doi = {10.1103/PhysRevLett.103.211101},
archivePrefix = {arXiv},
       eprint = {0908.1274},
 primaryClass = {astro-ph.SR},
       adsurl = {https://ui.adsabs.harvard.edu/abs/2009PhRvL.103u1101B}
}

@ARTICLE{Zhang2025,
       author = {{Zhang}, Michael F. and {Kunz}, Matthew W. and {Squire}, Jonathan and {Klein}, Kristopher G.},
        title = "{Extreme Heating of Minor Ions in Imbalanced Solar-wind Turbulence}",
      journal = {\apj},
         year = 2025,
        month = feb,
       volume = {979},
       number = {2},
          eid = {121},
        pages = {121},
          doi = {10.3847/1538-4357/ad95fc},
archivePrefix = {arXiv},
       eprint = {2408.04703},
 primaryClass = {astro-ph.SR},
       adsurl = {https://ui.adsabs.harvard.edu/abs/2025ApJ...979..121Z}
}

@ARTICLE{Villarroel2026,
       author = {{Villarroel-Sep{\'u}lveda}, Nicol{\'a}s and {Verscharen}, Daniel and {Moya}, Pablo S. and {L{\'o}pez}, Rodrigo A. and {Klein}, Kristopher G.},
        title = "{Wave─Particle Equilibria with Heavy Ions in Weakly Collisional Space Plasmas}",
      journal = {\apj},
         year = 2026,
        month = may,
       volume = {1002},
       number = {1},
          eid = {49},
        pages = {49},
          doi = {10.3847/1538-4357/ae5639},
archivePrefix = {arXiv},
       eprint = {2603.22613},
 primaryClass = {physics.plasm-ph},
       adsurl = {https://ui.adsabs.harvard.edu/abs/2026ApJ..1002...49V}
}

@ARTICLE{Walters2023,
       author = {{Walters}, Jada and {Klein}, Kristopher G. and {Lichko}, Emily and {Stevens}, Michael L. and {Verscharen}, Daniel and {Chandran}, Benjamin D.~G.},
        title = "{The Effects of Nonequilibrium Velocity Distributions on Alfv{\'e}n Ion-cyclotron Waves in the Solar Wind}",
      journal = {\apj},
         year = 2023,
        month = oct,
       volume = {955},
       number = {2},
          eid = {97},
        pages = {97},
          doi = {10.3847/1538-4357/acf1fa},
archivePrefix = {arXiv},
       eprint = {2308.14944},
 primaryClass = {astro-ph.SR},
       adsurl = {https://ui.adsabs.harvard.edu/abs/2023ApJ...955...97W}
}

@ARTICLE{Ran2026,
       author = {{Ran}, Hao and {Verscharen}, Daniel and {Coburn}, Jesse and {Nicolaou}, Georgios and {Ioannou}, Charalambos and {Wu}, Xiangyu and {Liu}, Jingting and {Klein}, Kristopher and {Owen}, Christopher},
        title = "{The Damping and Instability of Ion-acoustic Waves in the Solar Wind: Solar Orbiter Observations}",
      journal = {\apj},
         year = 2026,
        month = jun,
       volume = {1003},
       number = {2},
          eid = {124},
        pages = {124},
          doi = {10.3847/1538-4357/ae60fb},
archivePrefix = {arXiv},
       eprint = {2604.14311},
 primaryClass = {astro-ph.SR},
       adsurl = {https://ui.adsabs.harvard.edu/abs/2026ApJ..1003..124R}
}

@ARTICLE{Verscharen2018,
       author = {{Verscharen}, D. and {Klein}, K.~G. and {Chandran}, B.~D.~G. and {Stevens}, M.~L. and {Salem}, C.~S. and {Bale}, S.~D.},
        title = "{ALPS: the Arbitrary Linear Plasma Solver}",
      journal = {Journal of Plasma Physics},
         year = 2018,
        month = aug,
       volume = {84},
       number = {4},
          eid = {905840403},
        pages = {905840403},
          doi = {10.1017/S0022377818000739},
archivePrefix = {arXiv},
       eprint = {1803.04697},
 primaryClass = {physics.space-ph},
       adsurl = {https://ui.adsabs.harvard.edu/abs/2018JPlPh..84d9003V}
}

@ARTICLE{Klein2025,
       author = {{Klein}, K.~G. and {Verscharen}, D.},
        title = "{The dielectric response of plasmas with arbitrary gyrotropic velocity distributions}",
      journal = {Physics of Plasmas},
         year = 2025,
        month = sep,
       volume = {32},
       number = {9},
          eid = {092104},
        pages = {092104},
          doi = {10.1063/5.0286477},
archivePrefix = {arXiv},
       eprint = {2506.16431},
 primaryClass = {physics.plasm-ph},
       adsurl = {https://ui.adsabs.harvard.edu/abs/2025PhPl...32i2104K}
}

@ARTICLE{Opie2022,
       author = {{Opie}, Simon and {Verscharen}, Daniel and {Chen}, Christopher H.~K. and {Owen}, Christopher J. and {Isenberg}, Philip A.},
        title = "{Conditions for Proton Temperature Anisotropy to Drive Instabilities in the Solar Wind}",
      journal = {\apj},
         year = 2022,
        month = dec,
       volume = {941},
       number = {2},
          eid = {176},
        pages = {176},
          doi = {10.3847/1538-4357/ac982f},
archivePrefix = {arXiv},
       eprint = {2210.04875},
 primaryClass = {physics.space-ph},
       adsurl = {https://ui.adsabs.harvard.edu/abs/2022ApJ...941..176O}
}

@ARTICLE{Opie2024,
       author = {{Opie}, Simon and {Verscharen}, Daniel and {Chen}, Christopher H.~K. and {Owen}, Christopher J. and {Isenberg}, Philip A. and {Sorriso-Valvo}, Luca and {Franci}, Luca and {Matteini}, Lorenzo},
        title = "{Temperature anisotropy instabilities driven by intermittent velocity shears in the solar wind}",
      journal = {Journal of Plasma Physics},
         year = 2024,
        month = nov,
       volume = {90},
       number = {6},
          eid = {905900602},
        pages = {905900602},
          doi = {10.1017/S0022377824001375},
archivePrefix = {arXiv},
       eprint = {2409.18849},
 primaryClass = {physics.space-ph},
       adsurl = {https://ui.adsabs.harvard.edu/abs/2024JPlPh..90f9002O}
}

@ARTICLE{Klein2018,
       author = {{Klein}, K.~G. and {Alterman}, B.~L. and {Stevens}, M.~L. and {Vech}, D. and {Kasper}, J.~C.},
        title = "{Majority of Solar Wind Intervals Support Ion-Driven Instabilities}",
      journal = {\prl},
         year = 2018,
        month = may,
       volume = {120},
       number = {20},
          eid = {205102},
        pages = {205102},
          doi = {10.1103/PhysRevLett.120.205102},
archivePrefix = {arXiv},
       eprint = {1804.06330},
 primaryClass = {physics.space-ph},
       adsurl = {https://ui.adsabs.harvard.edu/abs/2018PhRvL.120t5102K}
}

@ARTICLE{Zhu2023,
       author = {{Zhu}, Xingyu and {Verscharen}, Daniel and {He}, Jiansen and {Maruca}, Bennett A. and {Owen}, Christopher J.},
        title = "{Regulation of Proton-{\ensuremath{\alpha}} Differential Flow by Compressive Fluctuations and Ion-scale Instabilities in the Solar Wind}",
      journal = {\apj},
         year = 2023,
        month = oct,
       volume = {956},
       number = {1},
          eid = {66},
        pages = {66},
          doi = {10.3847/1538-4357/aced03},
archivePrefix = {arXiv},
       eprint = {2308.02036},
 primaryClass = {physics.space-ph},
       adsurl = {https://ui.adsabs.harvard.edu/abs/2023ApJ...956...66Z}
}

@ARTICLE{Hellinger2019,
       author = {{Hellinger}, Petr and {Matteini}, Lorenzo and {Landi}, Simone and {Franci}, Luca and {Verdini}, Andrea and {Papini}, Emanuele},
        title = "{Turbulence versus Fire-hose Instabilities: 3D Hybrid Expanding Box Simulations}",
      journal = {\apj},
         year = 2019,
        month = oct,
       volume = {883},
       number = {2},
          eid = {178},
        pages = {178},
          doi = {10.3847/1538-4357/ab3e01},
archivePrefix = {arXiv},
       eprint = {1908.07760},
 primaryClass = {physics.space-ph},
       adsurl = {https://ui.adsabs.harvard.edu/abs/2019ApJ...883..178H}
}

@ARTICLE{Verscharen2016,
       author = {{Verscharen}, Daniel and {Chandran}, Benjamin D.~G. and {Klein}, Kristopher G. and {Quataert}, Eliot},
        title = "{Collisionless Isotropization of the Solar-wind Protons by Compressive Fluctuations and Plasma Instabilities}",
      journal = {\apj},
         year = 2016,
        month = nov,
       volume = {831},
       number = {2},
          eid = {128},
        pages = {128},
          doi = {10.3847/0004-637X/831/2/128},
archivePrefix = {arXiv},
       eprint = {1605.07143},
 primaryClass = {physics.space-ph},
       adsurl = {https://ui.adsabs.harvard.edu/abs/2016ApJ...831..128V}
}

@ARTICLE{Hellinger2015,
       author = {{Hellinger}, Petr and {Matteini}, Lorenzo and {Landi}, Simone and {Verdini}, Andrea and {Franci}, Luca and {Tr{\'a}vn{\'\i}{\v{c}}ek}, Pavel M.},
        title = "{Plasma Turbulence and Kinetic Instabilities at Ion Scales in the Expanding Solar Wind}",
      journal = {\apjl},
         year = 2015,
        month = oct,
       volume = {811},
       number = {2},
          eid = {L32},
        pages = {L32},
          doi = {10.1088/2041-8205/811/2/L32},
archivePrefix = {arXiv},
       eprint = {1508.03159},
 primaryClass = {physics.space-ph},
       adsurl = {https://ui.adsabs.harvard.edu/abs/2015ApJ...811L..32H}
}

@ARTICLE{Alexandrova2013,
       author = {{Alexandrova}, O. and {Chen}, C.~H.~K. and {Sorriso-Valvo}, L. and {Horbury}, T.~S. and {Bale}, S.~D.},
        title = "{Solar Wind Turbulence and the Role of Ion Instabilities}",
      journal = {\ssr},
         year = 2013,
        month = oct,
       volume = {178},
       number = {2-4},
        pages = {101-139},
          doi = {10.1007/s11214-013-0004-8},
archivePrefix = {arXiv},
       eprint = {1306.5336},
 primaryClass = {astro-ph.SR},
       adsurl = {https://ui.adsabs.harvard.edu/abs/2013SSRv..178..101A}
}

@ARTICLE{Bruno2013,
       author = {{Bruno}, Roberto and {Carbone}, Vincenzo},
        title = "{The Solar Wind as a Turbulence Laboratory}",
      journal = {Living Reviews in Solar Physics},
         year = 2013,
        month = dec,
       volume = {10},
       number = {1},
          eid = {2},
        pages = {2},
          doi = {10.12942/lrsp-2013-2},
       adsurl = {https://ui.adsabs.harvard.edu/abs/2013LRSP...10....2B}
}

@ARTICLE{Matteini2012,
       author = {{Matteini}, Lorenzo and {Hellinger}, Petr and {Landi}, Simone and {Tr{\'a}vn{\'\i}{\v{c}}ek}, Pavel M. and {Velli}, Marco},
        title = "{Ion Kinetics in the Solar Wind: Coupling Global Expansion to Local Microphysics}",
      journal = {\ssr},
         year = 2012,
        month = nov,
       volume = {172},
       number = {1-4},
        pages = {373-396},
          doi = {10.1007/s11214-011-9774-z},
       adsurl = {https://ui.adsabs.harvard.edu/abs/2012SSRv..172..373M}
}

@ARTICLE{Verscharen2022,
       author = {{Verscharen}, Daniel and {Chandran}, B.~D.~G. and {Boella}, E. and {Halekas}, J. and {Innocenti}, M.~E. and {Jagarlamudi}, V.~K. and {Micera}, A. and {Pierrard}, V. and {{\v{S}}tver{\'a}k}, {\v{S}}. and {Vasko}, I.~Y. and et al.},
        title = "{Electron-Driven Instabilities in the Solar Wind}",
      journal = {Frontiers in Astronomy and Space Sciences},
         year = 2022,
        month = aug,
       volume = {9},
          eid = {951628},
        pages = {951628},
          doi = {10.3389/fspas.2022.951628},
archivePrefix = {arXiv},
       eprint = {2206.10403},
 primaryClass = {physics.space-ph},
       adsurl = {https://ui.adsabs.harvard.edu/abs/2022FrASS...9.1628V}
}

@ARTICLE{Feldman1975,
       author = {{Feldman}, W.~C. and {Asbridge}, J.~R. and {Bame}, S.~J. and {Montgomery}, M.~D. and {Gary}, S.~P.},
        title = "{Solar wind electrons}",
      journal = {\jgr},
         year = 1975,
        month = nov,
       volume = {80},
       number = {31},
        pages = {4181},
          doi = {10.1029/JA080i031p04181},
       adsurl = {https://ui.adsabs.harvard.edu/abs/1975JGR....80.4181F}
}

@ARTICLE{Pilipp1987,
       author = {{Pilipp}, W.~G. and {Miggenrieder}, H. and {Montgomery}, M.~D. and {M{\"u}hlh{\"a}user}, K.-H. and {Rosenbauer}, H. and {Schwenn}, R.},
        title = "{Characteristics of electron velocity distribution functions in the solar wind derived from the helios plasma experiment}",
      journal = {\jgr},
         year = 1987,
        month = feb,
       volume = {92},
       number = {A2},
        pages = {1075-1092},
          doi = {10.1029/JA092iA02p01075},
       adsurl = {https://ui.adsabs.harvard.edu/abs/1987JGR....92.1075P}
}

@ARTICLE{Halekas2020,
       author = {{Halekas}, J.~S. and {Whittlesey}, P. and {Larson}, D.~E. and {McGinnis}, D. and {Maksimovic}, M. and {Berthomier}, M. and {Kasper}, J.~C. and {Case}, A.~W. and {Korreck}, K.~E. and {Stevens}, M.~L. and et al.},
        title = "{Electrons in the Young Solar Wind: First Results from the Parker Solar Probe}",
      journal = {\apjs},
         year = 2020,
        month = feb,
       volume = {246},
       number = {2},
          eid = {22},
        pages = {22},
          doi = {10.3847/1538-4365/ab4cec},
archivePrefix = {arXiv},
       eprint = {1912.02216},
 primaryClass = {astro-ph.SR},
       adsurl = {https://ui.adsabs.harvard.edu/abs/2020ApJS..246...22H}
}

@article{Navarro2020,
doi = {10.3847/2041-8213/aba0ae},
url = {https://doi.org/10.3847/2041-8213/aba0ae},
year = {2020},
month = {jul},
publisher = {The American Astronomical Society},
volume = {898},
number = {1},
pages = {L9},
author = {Navarro, Roberto E. and Muñoz, Víctor and Valdivia, Juan A. and Moya, Pablo S.},
title = {Feasibility of Ion-cyclotron Resonant Heating in the Solar Wind},
journal = {The Astrophysical Journal Letters}
}

@article{Wicks2016,
doi = {10.3847/0004-637X/819/1/6},
url = {https://doi.org/10.3847/0004-637X/819/1/6},
year = {2016},
month = {feb},
publisher = {The American Astronomical Society},
volume = {819},
number = {1},
pages = {6},
author = {Wicks, R. T. and Alexander, R. L. and Stevens, M. and III, L. B. Wilson and Moya, P. S. and Viñas, A. and Jian, L. K. and Roberts, D. A. and O’Modhrain, S. and Gilbert, J. A. and Zurbuchen, T. H.},
title = {A PROTON-CYCLOTRON WAVE STORM GENERATED BY UNSTABLE PROTON DISTRIBUTION FUNCTIONS IN THE SOLAR WIND},
journal = {The Astrophysical Journal}
}

@article{Moya2014,
doi = {10.1088/0004-637X/781/2/76},
url = {https://doi.org/10.1088/0004-637X/781/2/76},
year = {2014},
month = {jan},
publisher = {The American Astronomical Society},
volume = {781},
number = {2},
pages = {76},
author = {Moya, P. S. and Navarro, R. and Viñas, A. F. and Muñoz, V. and Valdivia, J. A.},
title = {WEAK TURBULENCE CASCADING EFFECTS IN THE ACCELERATION AND HEATING OF IONS IN THE SOLAR WIND},
journal = {The Astrophysical Journal}
}

@article{Jian2016,
    author = {Jian, L. K. and Moya, P. S. and Viñas, A. F. and Stevens, M.},
    title = {Electromagnetic cyclotron waves in the solar wind: Wind observation and wave dispersion analysis},
    journal = {AIP Conference Proceedings},
    volume = {1720},
    number = {1},
    pages = {040007},
    year = {2016},
    month = {03},
    issn = {0094-243X},
    doi = {10.1063/1.4943818},
    url = {https://doi.org/10.1063/1.4943818}
}

@article{Yoon2017,
  title = {Kinetic instabilities in the solar wind driven by temperature anisotropies},
  volume = {1},
  ISSN = {2367-3192},
  url = {http://dx.doi.org/10.1007/s41614-017-0006-1},
  DOI = {10.1007/s41614-017-0006-1},
  number = {1},
  journal = {Reviews of Modern Plasma Physics},
  publisher = {Springer Science and Business Media LLC},
  author = {Yoon,  Peter H.},
  year = {2017}
}

@BOOK{Stix:1992,
  title = {{Waves in plasmas}},
  year = {1992},
  author = {{Stix}, T.~H.},
  adsurl = {http://adsabs.harvard.edu/abs/1992wapl.book.....S},
  booktitle = {Waves in plasmas , by Stix, Thomas Howard.; Stix, Thomas Howard.~
	New York : American Institute of Physics, c1992.},
publisher = {American Institute of Physics}
}

@ARTICLE{Brown:2026,
       author = {{Brown}, Collin R. and {Howes}, Gregory G. and {Klein}, Kristopher G. and {TenBarge}, Jason M.},
        title = "{Phase-space energy transfer of wave─particle interactions using the field─particle correlation technique and linear plasma theory with JET-PLUME}",
      journal = {Physics of Plasmas},
         year = 2026,
        month = jul,
       volume = {33},
       number = {7},
          eid = {072101},
        pages = {072101},
          doi = {10.1063/5.0335826},
archivePrefix = {arXiv},
       eprint = {2606.20865},
 primaryClass = {physics.plasm-ph},
       adsurl = {https://ui.adsabs.harvard.edu/abs/2026PhPl...33g2101B}
}
\bibliographystyle{aasjournalv7}



\end{document}